\documentclass[preprintnumbers,nofootinbib,
superscriptaddress,amsmath,floatfix,prd]{revtex4}

\usepackage{amssymb}  
\usepackage{graphicx,subfigure}
\usepackage{exscale}
\usepackage{textcomp}
\usepackage{enumerate}
\usepackage{amsmath}
\usepackage{amssymb}
\usepackage[compat=1.1.0]{tikz-feynhand} 

\usepackage{color}

\usepackage[normalem]{ulem}  

\newcommand{\be}{\begin{equation}}
\newcommand{\ee}{\end{equation}}
\newcommand{\bea}{\begin{eqnarray}}
\newcommand{\eea}{\end{eqnarray}}
\newcommand{\ba}[1]{\begin{array}{#1}}
\newcommand{\ea}{\end{array}}

\newcommand{\bm}[1]{\mbox{\boldmath${#1}$}}

\newcommand{\Msolar}{\ensuremath{{\rm M}_\odot}}

\newcommand{\sliver}{\kern 0.07em} 

\begin{document}

\title{Strangeness-changing Rates and Hyperonic Bulk Viscosity in Neutron Star Mergers}
\author{Mark G.~Alford}
\email{alford@wustl.edu}

\author{Alexander Haber}
\email{ahaber@physics.wustl.edu}
\affiliation{Physics Department, Washington University in Saint Louis, 63130 Saint Louis, MO, USA}

\date{December 6, 2020}

\begin{abstract}
In this paper we present a computation of
the rates of strangeness-changing processes and the resultant
bulk viscosity in matter at the densities and temperatures typical of neutron star mergers. To deal with the high temperature in this environment we go beyond the Fermi surface approximation in our rate calculations and numerically evaluate the full phase space integral.  We include processes where quarks move between baryons via meson exchange: these have generally been  omitted in previous analyses but provide the dominant contribution to the rates of strangeness-changing processes and the bulk viscosity. The calculation of these rates is an essential step towards any calculation of dissipation mechanisms in hyperonic matter in mergers. As one application, we calculate the dissipation times for density oscillations at the frequencies seen in merger simulations. We find that hyperon bulk viscosity for temperatures in the MeV regime can probably be neglected in this context, but becomes highly relevant for keV-range temperatures. 
\end{abstract}

\maketitle

\section{Introduction}
\label{sec:intro}
The discovery of gravitational waves from a binary neutron star merger in 2017, named GW170817 \cite{TheLIGOScientific:2017qsa}, opened a
new window to study dense nuclear and possibly quark matter at high densities and temperatures \cite{Baiotti:2019sew,Raithel:2019uzi,Alford:2019oge,Chatziioannou:2019yko,Weih:2019xvw,Most:2019onn,Chatziioannou:2020pqz,Radice:2020ddv}. In order to relate the phase structure of dense matter to the astrophysical observations detailed simulations using numerical relativity and relativistic hydrodynamics have to be performed \cite{Baiotti:2016qnr, Hanauske:2017oxo,Radice:2017zta,Perego:2019adq,Hanauske:2019qgs},
using accurate representations of the relevant material properties. Therefore, it is necessary to improve our understanding of dense matter in merger conditions. Studies of GW170817 \cite{Abbott:2018exr,Landry:2018prl} estimate that the central densities of the merging stars were more than two times saturation density ($n_0=0.15\,\mathrm{fm}^{-3})$.  Numerical simulations of the first $20$\,ms after the initial contact of the stars provide further insight. They suggest that the density reaches several times saturation density and that temperatures can reach tens of MeV \cite{Hanauske:2017oxo,Hanauske:2019qgs}, where some simulations even predict up to $T\approx 100$\,MeV \cite{Perego:2019adq}. Furthermore, fluid elements undergo strong density oscillations with central frequencies of around $1$\,kHz \cite{Alford:2017rxf,Bernuzzi:2015opx}.  This raises the question of which microscopic transport phenomena and dissipation mechanisms are important on the $20$\,ms neutron star merger time scale. Initial estimates of various transport phenomena in Ref.~\cite{Alford:2017rxf} showed the potential importance of bulk viscosity in ordinary nuclear matter. Bulk viscosity is a dissipative mechanism, which converts oscillation energy into heat or radiated neutrinos. The magnitude of the bulk viscosity and the equation of state (EOS) of nuclear matter together determine the dissipation time scale on which oscillations are damped. A detailed study in neutrino-transparent matter showed that dissipation times for $npe$-matter due to direct and modified Urca processes are indeed on a millisecond timescale \cite{Alford:2019qtm,Harris:2020rus}, whereas in the neutrino-trapped regime, bulk viscosity seems to be negligible \cite{Alford:2019kdw,Alford:2020lla}. 

The intriguing prospect for nuclear physics is that other forms of matter might have different bulk viscosity, leading to observable signatures of their presence in the merger. 
In this paper we focus on hyperonic matter, where several weak, non-leptonic processes can contribute to beta equilibration and hence to bulk viscosity. Although the existence of hyperons in cold, isolated neutron stars is contested (the  ``hyperon puzzle'' \cite{Chatterjee:2015pua, Vidana:2018bdi}), the higher temperatures and densities reached in the merger render their appearance highly likely. In the past, hyperonic bulk viscosity has been exclusively studied at low (keV range) temperature, often in the context of r-modes \cite{Jones:2001ya,Haensel:2001em,Lindblom:2001hd,vanDalen:2003uy,Chatterjee:2006tk,Gusakov:2008hv,Haskell:2010ab, Ofengeim:2019fjy}. At these temperatures one can use the Fermi surface (FS) approximation since all particles participating in beta equilibration processes are close to their Fermi surfaces. Furthermore, an ultra non-relativistic approach, where the baryon momenta in the matrix element are set to zero, is sometimes adopted \cite{Haensel:2001em,Jones:2001ya, Haskell:2010ab}  in order to obtain analytic results. In the merger environment, both of these assumptions are invalid and need to be improved on.  Additionally, most studies only consider the contact interaction diagram where a $W$ boson is exchanged between baryons. In Refs.~\cite{vanDalen:2003uy,Ofengeim:2019fjy}, it has been shown that, at least at the studied low temperatures, the one meson exchange (OME) contribution, where the $W$ exchange is internal to a hadron, dominates the rates that are relevant to the bulk viscosity. 
In our treatment of the beta equilibration rate we improve on previous treatments and obtain results that are valid in the merger environment by\\
(a) Taking the OME contributions for all processes into account; \\
(b) Computing numerically the full twelve dimensional phase space integral instead of using the FS approximation; \\
(c) Using a fully relativistic approach, which is particularly important at high densities where the Fermi momenta are largest. 

This allows us to calculate the re-equilibration rates for four different strangeness changing, weak decay processes, two of which predominantly occur via OME, not via the contact interaction which is heavily suppressed. We show that all of these rates contribute to the bulk viscosity and have to be taken into account. We find that the re-equilibration rates at high temperatures are generally too fast to lead to a sizeable bulk viscosity and correspondingly short dissipation times. Consequently, hyperon bulk viscosity is most likely not a significant factor in the hot regions of neutron star mergers. However, for temperatures in the keV regime, bulk viscosity shows a resonant peak, giving damping times in the ms range, which means that we expect significant damping of density oscillations in either exceptionally cold regions of a merger or in the inspiral phase of an elliptic merger, where tidal forces are expected to excite f-modes with frequencies of order 1\,kHz. \cite{Chirenti:2016xys,Pratten:2019sed}.
The processes whose rates we calculate, along with other (semi-leptonic) hyperon decay processes, might play an important role in  cooling, thermal transport or radiative dissipation \cite{Sad:2009hba} and are a fundamental ingredient for an extension of the calculation of phase conversion dissipation \cite{Alford:2014jha} to merger temperatures.

In this paper we use natural units, where $\hbar=c=k_B=1$ and the mostly-minus signature of the Minkowski metric, $g^{\mu\nu}=\mathrm{diag}(1,-1,-1,-1)$.

\section{Hyperonic Matter and Bulk Viscosity}
\label{sec:hyperonic}
\subsection{Equation of State}
There are many proposed equations of state EOS for nuclear matter with hyperonic degrees of freedom. Depending on the EOS, different hyperons appear at different onset densities \cite{Schaffner:1995th,Weber:2004kj,SchaffnerBielich:2008kb,Colucci:2013pya,vanDalen:2014mqa,Oertel:2015fta,Li:2018jvz,Spinella:2018dab}. Since our analysis requires calculations, including derivatives, of the EOS both in and out of chemical equilibrium with respect to strangeness, we use a simple EOS, that we call ``PK1+H'', which can be computed at arbitrary strangeness fraction, rather than using an EOS that is defined via a table of numbers. PK1+H allows stars up to a maximum mass of 1.88\,\Msolar, putting it at the $3\sigma$ edge of compatibility with current constraints ($M_{\rm max}\geqslant 1.928\pm 0.017\,\Msolar$ \cite{Fonseca:2016tux}, $M_{\rm max}\geqslant 2.01\pm 0.04\,\Msolar$ \cite{Antoniadis:2013pzd}). To check that our main conclusions are not specific to the EOS that we used, we computed the peak bulk viscosity using another hyperonic EOS, GM1'B, which has $M_{\rm max}= 2.02\,\Msolar$ \cite{Gusakov:2014ota}.  GM1'B predicts a different order of the onset of the different hyperon species and includes an additional (strange) exchange meson, which leads to a repulsion between the hyperons. However, as we will discuss in Sec.~\ref{sec:results}, in the relevant density and temperature range it predicts a maximum bulk viscosity comparable to  PK1+H.  This is an indication that our findings concerning the relevance of hyperonic bulk viscosity are valid for any EOS where at least one hyperonic degree of freedom appears at a density that is reachable in mergers. 

The PK1+H EOS is based on a relativistic mean field model (RMF) which includes nonlinear mesonic terms which interact with the nucleons and the $\Lambda$ and $\Sigma^-$ hyperons, which have the lowest onset densities. We neglect the other hyperons in the baryon octet because they only appear at much higher densities. In PK1+H, the  $\Sigma^-$ hyperon appears first as a function of density due to its contributions to the overall charge neutrality of matter. The nuclear part of the Lagrangian including the Yukawa couplings $g_{\sigma N},\ g_{\omega N},\ g_{\rho N}$, between the nucleons and the three mesons follow the conventions in \cite{SUGAHARA1994557}, the numerical parameters are chosen according to the PK1 parametrization from Table~1 in Ref.~\cite{Long:2003dn}. We extend the PK1 EOS to the hyperonic sector by adding the hyperons to the Langrangian as shown below.  The hyperonic coupling constants are chosen in accordance with Ref.~\cite{Schaffner:1995th} in such a way that the model reproduces a hyperon spectrum similar to the one from the DD-ME2 hyperonic EOS investigated in Ref.~\cite{Colucci:2013pya}. All numerical parameters are summarized in App.~\ref{app:const}. 
The Lagrangian of the model is
\begin{subequations}
\begin{align}
\mathcal{L}&=\mathcal{L}_B+\mathcal{L}_m+\mathcal{L}_l\, ,\\[2ex]
\mathcal{L}_B&=\sum_i \bar{\psi}_i\left[ i\gamma^{\mu}\partial_{\mu}-M_i-g_{\sigma i}\sigma-g_{\omega i} \gamma^{\mu}\omega_{\mu}-g_{\rho i}\gamma^{\mu} \vec{\tau}\cdot\vec{\rho}_{\mu}\right]\psi_i \, ,\\[2ex]
\mathcal{L}_m &=\frac{1}{2} \partial_\mu\sigma\partial^\mu\sigma-\frac{1}{2}m_\sigma^2\sigma^2-\frac{g_2}{3}\sigma^3-\frac{g_3}{4}\sigma^4-\frac{1}{4}\omega^{\mu\nu}\omega_{\mu\nu}+\frac{1}{2}m_\omega^2\omega^\mu\omega_\mu+\frac{c_3}{4}\left(\omega^\mu\omega_\mu\right)^2+\frac{1}{2}m_\rho^2\vec{\rho}^{\,\mu}\cdot\vec{\rho}_\mu-\frac{1}{4}\vec{\rho}^{\,\mu\nu}\cdot\vec{\rho}_{\mu\nu}\, ,\\[2ex]
\mathcal{L}_l&=\sum_l\bar{\psi}_l\left[ i\gamma^{\mu}\partial_{\mu}-m_l\right]\psi_l \, ,
\end{align}
\end{subequations}
where
\begin{subequations}
\begin{align}
  \omega^{\mu\nu} &\equiv \partial^\mu\omega^\nu-\partial^\nu\omega^\mu \, ,  \\[2ex]
  \vec{\rho}^{\,\mu\nu}&\equiv \partial^\mu\vec{\rho}^{\,\nu}-\partial^\nu\vec{\rho}^{\,\mu}+g_{\rho N}\vec{\rho}^{\,\mu}\times\vec{\rho}^{\,\nu}\, ,
\end{align}
\end{subequations}
with symbols with arrows being vectors in isospin space. The first term $\mathcal{L}_B$ includes the sum over the four baryons (neutron, proton, $\Lambda$ and $\Sigma^{-}$) with their masses $M_i$ and their Yukawa interactions with the mesonic fields. We include the scalar $\sigma$ meson, the vector meson $\omega^\mu$ and the isovector triplet $\vec{\rho}^{\,\mu}$, which breaks isospin symmetry, and self-interactions for the scalar and the vector mesons. Note that the Yukawa couplings are different for every baryon-meson interaction. Their values are given in App.~\ref{app:const}. The leptonic Lagrangian $\mathcal{L}_l$ introduces free electrons and muons, where we assume the electrons to be massless. The particle fractions in or out of chemical equilibrium are then obtained by solving the Euler-Lagrange equations  in the mean field approximation.  
\begin{figure} [t]
\begin{center}
\includegraphics[width=0.5\textwidth]{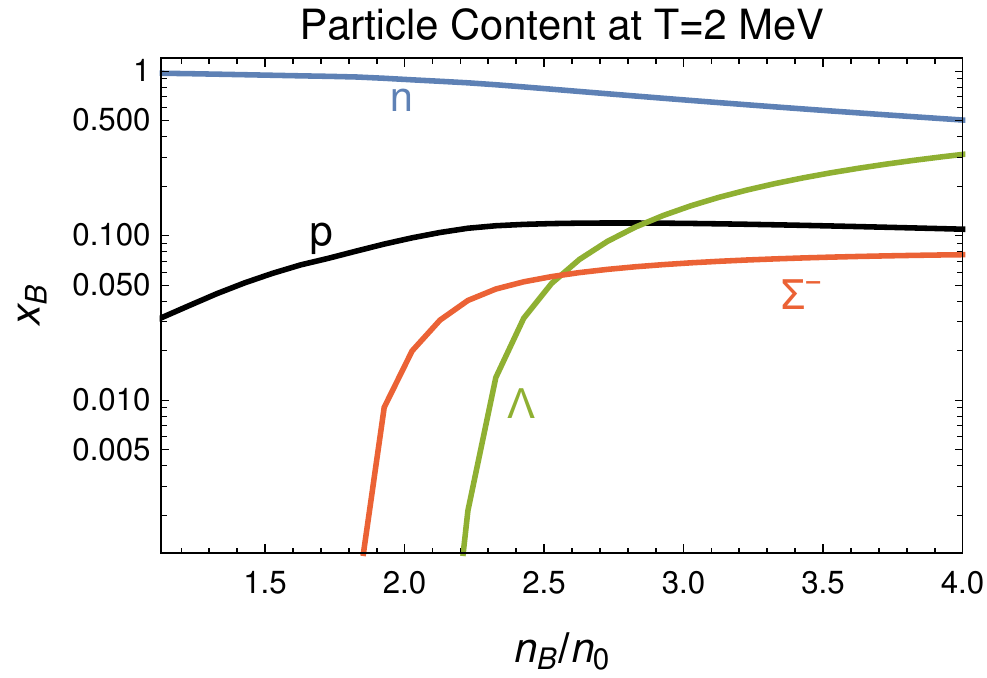}
\caption{Logarithm of the ratios of all baryonic particle densities over the total baryon density at $T=2$\,MeV plotted as a function of total baryon density in units of the saturation density. In the used parametrization PK1+H, saturation density is given by $n_0=0.148~\mathrm{fm}^{-3}$. Although the $\Lambda$-hyperon is less massive than the $\Sigma^-$hyperon, the order of their onset is reversed because of charge neutrality.}
\label{fig:fracs}
\end{center}
\end{figure}  

The resultant particle content for neutral matter in beta equilibrium is shown in Fig.~\ref{fig:fracs}. Chemical equilibrium, charge neutrality and baryon number can be expressed as 
\begin{subequations}
\begin{align}
  n_B&= n_n + n_p +n_{\Sigma^-}+n_\Lambda & &\mathrm{baryon\ number} \, ,  \\[2ex]
   n_p&= n_e + n_\mu +n_{\Sigma^-} & &\mathrm{charge\ neutrality} \, ,  \\[2ex]
   \mu_p&=\mu_n-\mu_e & &\mathrm{chemical\ equilibrium} \, , \\[2ex]
    \mu_e&=\mu_{\mu}  \, , \\[2ex]
    \mu_{\Sigma^-}&=\mu_n+\mu_e \, , \\[2ex]
    \mu_\Lambda &= \mu_n \, ,
\end{align}
\end{subequations}
where $n_i$ and $\mu_i$ are the number density and chemical potential for particle species $i$.
The resulting dispersion relations for the baryons are given by
\begin{equation}
    E_i=\sqrt{p_i^2+\left(M_i^*\right)^2}+g_{\omega i}\langle\omega_0\rangle+g_{\rho i}I_{i3}\langle\rho_{03}\rangle \, ,
    \label{eq:fulldisp}
\end{equation}
with the modulus of the three-momentum $p_i=|\mathbf{p_i}|$, the effective baryon mass $M_i^*=M_i-g_{\sigma i}\langle\sigma\rangle$, where $\langle\sigma\rangle$ is the vacuum-expectation value (vev) of the $\sigma$-meson and $\langle\omega_0\rangle$ the vev of the temporal component of the $\omega$. Only the temporal part of the third isospin-vector component of the $\vec{\rho}$ develops a finite expectation value $I_{i3}\langle\rho_{03}\rangle$, where $I_{i3}$ denotes the third component of the isospin projection of the $i-$th baryon.

\subsection{Rate Calculation and Matrix Element}
Computations of hyperonic bulk viscosity have been performed using various nucleonic interactions, approximations and EOS in the past, but exclusively for low enough temperatures so that the FS approximation is valid, and often in the context of the r-mode instability \cite{Jones:2001ya,Lindblom:2001hd,vanDalen:2003uy,Chatterjee:2006hy, Haskell:2010ab, Ofengeim:2019fjy}. In this work we are interested in mergers where the temperature is high enough to eliminate  nucleonic or hyperonic superfluidity and to invalidate the FS approximation.

Hyperonic bulk viscosity arises from beta equilibration of the strangeness fraction, which will be dominated by the fastest strangeness-changing processes.
We focus on non-leptonic processes, which are typically faster than (semi-)leptonic ones \cite{vanDalen:2003uy,Alford:2010jf}.  The processes we are including in this work all change strangeness by one unit and are mediated by the weak interaction,
\begin{subequations}
\label{eq:allpr}
\begin{align}
  \text{I}:\ n+n &\Longleftrightarrow p + \Sigma^- \, , \label{eq:pr1}\\[2ex]
  \text{II}:\ n+p &\Longleftrightarrow p + \Lambda \, , \label{eq:pr2}\\[2ex]
   \text{III}:\ n+n &\Longleftrightarrow n + \Lambda \, , \label{eq:pr3}\\[2ex]
   \text{IV}:\ \Lambda+\Lambda &\Longleftrightarrow \Lambda + n \, \label{eq:pr4}.
\end{align}
\end{subequations}

\begin{figure}[t]
\begin{center}
\begin{tabular}{c@{\hspace{4em}}c@{\hspace{4em}}c}
\begin{tikzpicture}
\begin{feynhand}
\vertex [ringdot] (a) at (0,0) {} ;
\vertex [dot] (b) at (0,-2) {};
\node at (0.1,0.4) {F$^W_{n\Sigma}$};
\node at (0.1,-2.4) {F$^S_{np}$};
\vertex [particle] (c) at (-2,0) {n};
\vertex [particle] (d) at (2,0) {$\Sigma^-$};
\vertex [particle] (i3) at (-2,-2) {n};
\vertex [particle] (i4) at (2,-2) {p};
\propagator[plain] (c) to (a);
\propagator[plain] (a) to (d);
\propagator [scalar] (a) to [edge label=$\pi^-$](b);
\propagator[plain] (i3) to (b);
\propagator[plain] (b) to (i4);
\end{feynhand}
\end{tikzpicture}
&
\begin{tikzpicture}[scale=.5]
\begin{feynhand}
\vertex [particle] (o1) at (4,0) ;
\vertex [particle] (o2) at (4,-1) ;
\vertex [ringdot] (W1) at (3,-1) ;
\vertex [ringdot] (W2) at (3,-2) ;
\vertex [dot](o3) at (4,-2) ;
\vertex [dot] (u1) at (4,-4) ;
\vertex [particle](u2) at (4,-5) ;
\vertex [particle] (u3) at (4,-6) ;
\vertex [particle] (q1) at (0,0) {d};
\vertex [particle] (q2) at (0,-1) {u};
\vertex [particle] (q3) at (0,-2) {d};
\vertex [particle] (q4) at (0,-4) {d};
\vertex [particle] (q5) at (0,-5) {d};
\vertex [particle] (q6) at (0,-6) {u};

\vertex [particle] (q1p) at (8,0) {d};
\vertex [particle] (q2p) at (8,-1) {s};
\vertex [particle] (q3p) at (8,-2) {d};
\vertex [particle] (q4p) at (8,-4) {u};
\vertex [particle] (q5p) at (8,-5) {d};
\vertex [particle] (q6p) at (8,-6) {u};

\propagator[fermion] (q1) to (o1);
\propagator[fermion] (q2) to (W1);
\propagator[plain] (W1) to  [edge label =$s$] (o2);
\propagator[fermion] (q3) to  (W2);
\propagator[plain] (W2) to [edge label =$u$](o3);

\propagator[chabos] (W1) to [edge label'=$W^+$] (W2);

\propagator[fermion] (q4) to (u1);
\propagator[fermion] (q5) to (u2);
\propagator[fermion] (q6) to (u3);

\propagator[fermion] (o1) to (q1p);
\propagator[fermion] (o2) to (q2p);
\propagator[fermion] (o3) to (q3p);

\propagator[fermion] (u1) to (q4p);
\propagator[fermion] (u2) to (q5p);
\propagator[fermion] (u3) to (q6p);

\propagator[anti charged scalar] (o3) to (u1) ;

\node at (5.4,-3.0)  {$\pi^-=\bar{u}d$};
\node at (-.8,-1) {n};
\node at (-.8,-5) {n};
\node at (8.8,-1) {$\Sigma^-$};
\node at (8.8,-5) {p};

\node at (4.8,-1.6) {d};
\node at (4.8,-3.6) {u};

\end{feynhand}
\end{tikzpicture}
&
\begin{tikzpicture}[scale=.5]
\begin{feynhand}

\vertex [particle] (lu) at (-2,-2) {n};
\vertex [particle] (ro) at (2,2) {$\Sigma^-$};
\vertex [particle] (lo) at (-2,2) {n};
\vertex [particle] (ru) at (2,-2) {p};
\propagator[plain] (lu) to (ro);
\propagator[plain] (ru) to (lo);

\end{feynhand}
\end{tikzpicture}
\\
(a) & (b) & (c)
\end{tabular}

\caption{ Panels (a) and (b) show Feynman and quarkflow diagrams for the OME contribution to process I. 
The flavor changing weak-interaction vertex $F^W_{n\Sigma}$ connecting the incoming neutron $n$ with a pion and the $\Sigma^-$-hyperon represents a combination of a flavor changing $W-$boson exchange within the baryon and a quark exchange (modeled via one meson exchange) with the spectator baryon. The strong-interaction vertex F$^S_{np}$ connects the nucleons $n$ and $p$ with a pion. For the matrix element in Eq.~(\ref{eq:matelem}), we have to subtract a second Feynman diagram with the two initial neutrons exchanged. Panel (c) shows the Feynman diagram for the contact interaction contribution, where the two nucleons exchange a charged $W$-boson that is integrated out. This is the basis for the matrix element in Eq.~(\ref{eq:MnnpsC}). All coupling constants can be found in App.~\ref{app:const}. The remaining diagrams are shown in App.~\ref{app:dia}.}
\label{fig:_feyndia}
\end{center}
\end{figure}
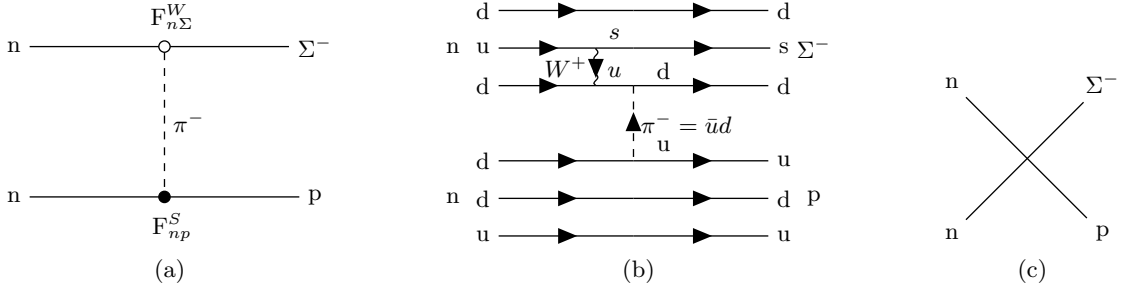

In general there are two main contributions to such processes.\\
(a) ``contact interaction'': exchange of a $W$ boson between
the baryons, which at the energy scales relevant to our calculations can be reduced to a contact interaction between the baryons, depicted for process I in Fig.~\ref{fig:_feyndia}(c);\\
(b) ``one meson exchange'' (OME): a combined weak-strong channel, depicted for process I in Fig.~\ref{fig:_feyndia}(a) and (b). In this channel, the flavor-changing W-boson exchange occurs inside one of the incoming baryons, creating an off-shell intermediate state. A strong interaction with the second incoming baryon rearranges the quarks and improves the kinematics of the process. We model that strong interaction as one-meson exchange.

Early work by Jones \cite{Jones:2001ya} and Lindblom and Owen \cite{Lindblom:2001hd} only included contact interactions, so they neglected processes III and IV which
would require exchange of a Z boson between the baryons, and such flavor changing neutral currents are highly suppressed by the GIM mechanism \cite{1970PhRvD...2.1285G}.

However, there are OME contributions to all four processes in Eq.~(\ref{eq:allpr}), and at temperatures in the sub-MeV range 
the OME channel is the dominant contribution.
For processes I and II, the OME contribution to the rate is an order of magnitude larger than the contact interaction contribution \cite{vanDalen:2003uy}. Process III, in particular, is non-negligible at most densities. This can partially be attributed to the large phase space near the neutron Fermi surface compared to the other baryon species. 
We calculate the OME contribution to all 4 processes. The rates $\Gamma_{12\to 34}$ can be calculated either in the FS approximation for low temperatures, or by computing the full phase space integral:

\begin{align}  
    \Gamma_{12\to 34}= &\frac{1}{S}\int\frac{d^3p_1}{(2\pi)^3}\frac{d^3p_2}{(2\pi)^3}\frac{d^3p_3}{(2\pi)^3}\frac{d^3p_4}{(2\pi)^3}\frac{\sum_{\mathrm{s}}|M_{1234}|^2}{2^4E_1^*E_2^*E_3^*E_4^*}(2\pi)^4\delta\left(E_1+E_2-E_3-E_4\right)\delta^3\left(\mathbf{p_1}+\mathbf{p_2}-\mathbf{p_3}-\mathbf{p_4} \right)\times \label{eq:rateint} \\[2ex]
   &f_1(E_1,\mu_1) f_2(E_2,\mu_2) \left[1-f_3(E_3,\mu_3)\right]\left[1-f_4(E_4,\mu_4)\right] \, , \nonumber
\end{align}  
with the symmetry factor $S=2$ for all processes with two identical baryons on one side of the reaction, i.e.~processes I, III and IV, and $S=1$ for process II. The spin-summed, squared matrix element of the process $\sum_s|M_{1234}|^2$
turning the incoming baryons with labels $1$ and $2$ into baryons $3$ and $4$, where the labels stand for the corresponding baryons in Eqs.~(\ref{eq:allpr}),
can be obtained from the Feynman diagrams in panel (a) of Fig.~\ref{fig:_feyndia}
which give the matrix element
\be
M_{1234}=\left[\bar{u}_3F^S_{23}u_2\,\bar{u}_4F_{14}^Wu_1\,D_{\varphi}(k_1^2)-\bar{u_3}F^S_{13}u_1\,\bar{u}_4F_{24}^Wu_2\,D_{\varphi}(k_2^2) \right] \, , \label{eq:matelem}
\ee
where the Dirac bispinors are normalized following Refs.~\cite{griffiths2008introduction} and \cite{Roberts:2016mwj} to $u^{\dagger} u=2E^*$ which leads to the corresponding energy denominators in Eq.~(\ref{eq:rateint}).  When we evaluate $|M_{1234}|^2$,
the spin summation over Dirac bispinors, which follow equations of motion derived from meson exchange Lagrangians as used here, leads to an
expression in terms of the quasi-momentum $(E^*,-\bm{p})$ 
where
\begin{equation}
    E_i^*=\sqrt{p_i^2+\left(M_i^*\right)^2} \, .
\end{equation}
In all other parts of the calculation, including the delta distributions in the rate integral Eq.~(\ref{eq:rateint}),
on-shell nucleons are characterized by four-momenta that obey the dispersion relation Eq.~(\ref{eq:fulldisp}).
Therefore, the meson propagator $D_\varphi$, defined in Eq.~(\ref{eq:mesprop}), depends on the dispersion relations from Eq.~(\ref{eq:fulldisp}) as well, whereas the remaining matrix element is given in terms of the quasi-momentum. For a detailed calculation of spin sums in RMFs see appendix B of Ref.~\cite{Roberts:2016mwj}.

The weak and strong interaction vertices are given by
\begin{align}\label{eq:vertex}
    F_{ij}^W= G_F m_\pi^2\left(A_{ij}+B_{ij}\gamma_5 \right)\, , \qquad F_{ij}^S=g_{ij}\gamma_5\, ,
\end{align}
with the Fermi constant $G_F=1.1663787\times 10^{-5}$ GeV$^{-2}$, the fifth gamma matrix $\gamma_5$, and the strong interaction coupling constants $g_{ij}$ and weak interaction coupling constants $A_{ij}$ and $B_{ij}$, which depend on the baryons in the corresponding vertex and are summarized in App.~\ref{app:const}. The coupling constants $A_{ij}$ and $B_{ij}$ are rendered massless via the insertion of a factor of the pion mass squared, $m_{\pi}^2$, for all processes (whether the exchanged meson is a pion or not).  The meson propagator is given by
\be \label{eq:mesprop}
D_\varphi(k)=\frac{1}{k_0^2-k^2-m_{\varphi}^2} \, ,
\ee
where the energy $k_0$ and the momentum $k$ of the meson $\varphi$, which would be a pion in processes I to III and a kaon in IV, 
is determined by energy-momentum conservation in the vertices.

The Fermi-Dirac distribution functions
\begin{equation}
    f_i(E_i,\mu_i)=\frac{1}{1+\exp\left(\frac{E_i-\mu_i}{T}\right)}
\end{equation}
account for Pauli blocking and depend on the full dispersion relation of the incoming ($i=1,2$) and outgoing ($i=3,4$) baryons, see Eq.~(\ref{eq:fulldisp}), the chemical potentials $\mu_i$ and the temperature $T$. Since the effective masses become smaller than the corresponding Fermi momenta at high densities, we treat all baryons as relativistic particles. A non-relativistic treatment leads to nonphysical behavior of the bulk viscosity at medium to high densities (around $n_B\approx3n_0$) \cite{Alford:2020lla}. 

Although they will turn out to be small compared to the OME channel, we also compute the rates for the processes $n+n\leftrightarrow p + \Sigma^-$ and $n+p\leftrightarrow p + \Lambda$ in the contact interaction channel. The corresponding matrix elements are derived from the Feynman diagrams in Fig.~\ref{fig:_feyndia}(c), Fig.~\ref{Fig:quarkIIa}(c) and Fig.~\ref{Fig:quarkIIb}(c) and are, after spin-summation, given by \cite{Haensel:2001em,Lindblom:2001hd,vanDalen:2003uy,Ofengeim:2019fjy} 
\be\label{eq:MnnpsC}
    \sum_s\left|M_{nnp\Sigma^-} \right|^2=8G_F^2\sin^2(2\theta_C)M_n^2M_pM_{\Sigma^-}\left(1+3c_A^{np}c_A^{n\Sigma^-}\right)^2 
\ee
and
\be\label{eq:MnpplC}
   \sum_s\left|M_{npp\Lambda} \right|^2=8G_F^2\sin^2(2\theta_C)M_nM_p^2M_{\Lambda}\left(1+3|c_A^{np}|^2|c_A^{p\Lambda}|^2\right) \, . 
\ee
All numerical constants can be found in App.~\ref{app:const}. Since the OME processes provide the dominant contribution to the rates, we only need to make a rough estimate of the subdominant contribution from contact interactions. Following the approach used widely in the literature \cite{Haensel:2001em,Jones:2001ya, Haskell:2010ab}
we simplify the matrix element by applying the ultra non-relativistic approximation, where\ $E_i=M_i^*$ 
and the energy denominators in Eq.~(\ref{eq:rateint}) are replaced with the effective masses $M_i^*$. The contact interaction contribution to the rates can then  be computed analytically. We emphasize
that this is an extremely crude approximation: in cold hyperonic matter, the ultra non-relativistic approximation underestimates the rates by up to two orders of magnitude. However, even in an improved relativistic treatment, the contact interaction produces rates which are still an order of magnitude slower than the ones derived from the OME process \cite{vanDalen:2003uy,Ofengeim:2019fjy}.

The results for the rates in the OME and contact interaction channel are shown in Sec.~\ref{sec:results}. At low temperatures, the Fermi spheres are sharply defined and only particles close to the Fermi surface can participate in the processes given in Eqs.~(\ref{eq:allpr}). In this case, we can simplify the full phase space integral from Eq.~(\ref{eq:rateint}) by using the FS approximation: we fix all the momentum magnitudes to their respective Fermi momenta, and split the integral into  angular and energy contributions. The FS approximation can be applied to the OME and contact-interaction contributions. For details on the FS approximation see Refs.~\cite{Yakovlev:2000jp,Kaminker:2016ayg}. For a momentum-independent matrix element, like the contact interaction channel matrix element in the ultra non-relativistic approximation, the rate is
\be
\Gamma_{12\to 34}=\frac{T^{3}|M_{1234}|^{2}}{(2\pi)^{5}2^{3}S}I(\xi)Q^{(4)}\, ,\quad \mathrm{where} \quad I(\xi)= \frac{e^{\xi}}{e^{\xi}-1}\frac{4\pi^{2}\xi+\xi^{3}}{6} \, ,
\ee
and the squared matrix element $|M_{1234}|^2$ comes from Eq.~(\ref{eq:MnnpsC}) and Eq.~(\ref{eq:MnpplC}), the symmetry factor S and where  $\xi\equiv\delta\mu/T$, and $\delta\mu$ is the chemical potential that measures the deviation from chemical equilibrium (Eq.~(\ref{eq:muD})). $Q^{(4)}$ depends on the relations of the various Fermi momenta and is defined in Tab.~1 of Ref.~\cite{Kaminker:2016ayg}.\footnote{In Ref.~\cite{Kaminker:2016ayg}, the baryons are ordered by the magnitude of their Fermi momenta.} For the full momentum dependent matrix element from Fig.~\ref{fig:_feyndia}, the rate in the FS approximation  is
\be
\Gamma_{12\to 34}=\frac{M_1^*M_2^*M_3^*M_4^*}{S(2\pi)^82^4}p_{F4}T^3I(\xi)\int_0^{2\pi} d\varphi \int_{s_-}^{s_+} ds\frac{k_{1}^+|M_{1234}|^2(k_1^+)\theta(r_+^2-1)+k_{1}^-|M_{1234}|^2(k_1^-)\theta(r_+^2-1)}{\sqrt{p_{F2}^2-(1-s^2)p_{F4}^2}} \, ,\label{eq:rateFS}
\ee
where $p_{Fi}$ is the Fermi momentum of the $i-$th particle in Fig.~\ref{fig:_feyndia}, with momentum transfers $\mathbf{k}_1=\mathbf{p}_1-\mathbf{p}_3$ and  $\mathbf{k}_2=\mathbf{p}_4-\mathbf{p}_2$. Energy-momentum conservation demands that the  moduli of the momentum transfer vectors are equal, $k_1=k_2$. Futhermore, the delta distribution has two zeros, which lead to the two separate contributions to the rate integral with the modulus for $\mathbf{k}_1$ (and therefore $\mathbf{k}_2$) given by $k_1^{\pm}=p_{F4}s\pm\sqrt{p_{F4}^2(s^2-1)+p_{F2}^2}$, and $\theta$ being the Heaviside function of $r_{\pm}=\left(p_{F1}^2-k_1^{\pm2}-p_{F3}^2\right)/\left( 2k_1^{\pm}p_{F3} \right)$. The angles $\varphi$ and $s=\cos\theta$ are the azimuthal and polar angle between $\mathbf{p}_4$ and $\mathbf{k}_1$. The integration boundaries for $s\in[-1,1]$ have to be chosen such that $k_1^{\pm}$ is real and positive. $|M|^2_{1234}(k_1^{\pm})$ is the spin summed, squared matrix element evaluated on the Fermi surface, i.e. $|\mathbf{p}_3|=p_{F3}$, $|\mathbf{p}_4|=p_{F4}$ and $|\mathbf{k}_1|=|\mathbf{k}_2|=k_1^{\pm}$. Energy momentum conservation sets the polar angle $\cos\theta_3\equiv r$ between $\mathbf{k}_1$ and $\mathbf{p}_3$ to $r=r_{\pm}$ defined above. Note that in chemical equilibrium, $\xi=0$ and $\lim_{\xi\to 0}I(\xi)=2\pi^2/3$. The remaining integrals are evaluated numerically.

\subsection{Bulk Viscosity} 
In this section we derive an expression for hyperonic bulk viscosity, where we use the methods and notation of Refs.~\cite{Gusakov:2008hv,Ofengeim:2019fjy}. As noted in Sec.~\ref{sec:intro}, these works calculated the equilibration rate assuming low temperatures characteristic of isolated neutron stars. In this work we obtain results which are valid at the densities and temperatures that arise in mergers, where a thermal population of hyperons below the zero temperature onset is present. 
Bulk viscosity is defined by the deviation from the equilibrium pressure $P_{0}$ via 
\begin{equation}\label{eq:bulkdef}
P-P_{0}=-\zeta\nabla\cdot\boldsymbol{v}\,,
\end{equation}
where $P$ is the current pressure of the fluid element which undergoes a harmonic oscillation of the form \mbox{$n=n_{0}+\delta n\exp(-i\omega t)$}, with the external oscillation frequency $\omega$, the equilibrium density $n_0$ and the amplitude of the oscillation $\delta n\ll n_0$. The bulk viscosity $\zeta$ is given by the coefficient of the divergence of the fluid element velocity $\boldsymbol{v}$ on the right hand side of Eq.~(\ref{eq:bulkdef}). The oscillation will push the matter out of beta equilibrium, which causes a difference $\Delta\Gamma$ between the backward and forward rates of the individual processes in Eqs.~(\ref{eq:allpr}).
In principle, the pressure $P$ is a function of the six particle species
numbers (proton, neutron, $\Lambda$ and $\Sigma^{-}-$hyperons ,
electrons and muons). We assume that thermal conduction operates fast enough that
the temperature is time-independent, so we calculate isothermal susceptibilities and bulk viscosity. For sufficiently low temperatures and long
wavelength of the density oscillations an adiabatic calculation might be appropriate~\cite{Alford:2019qtm}, but we will reserve the exploration of this regime for future work. Not all of the particle densities are independent, since the total baryon number is given by
\begin{equation}
n_{B}=n_{n}+n_{p}+n_{\Lambda}+n_{\Sigma^{-}}\,,
\end{equation}
 and we assume local charge neutrality,
\begin{equation}
n_{p}=n_{e}+n_{\mu}+n_{\Sigma^{-}}\,.
\end{equation}
We assume that strong interaction processes such as
\begin{equation}
n+\Lambda\Longleftrightarrow p+\Sigma^{-}\,,\label{eq:fast}
\end{equation}
are in equilibrium. It is important to note that this strong interaction, though it conserves strangeness, changes the neutron number. This means that choosing the neutron number as the equilibrating quantity (as done in Refs.~\cite{Lindblom:2001hd,vanDalen:2003uy} for
instance) is misleading \cite{Gusakov:2014ota,Jones:2001ya}.
We will calculate the bulk viscosity arising from the equilbration of the strangeness fraction, which is given by
$x_H=n_H/n_B$ with the hyperon number $n_H=n_{\Sigma^-}+n_{\Lambda}$ \cite{Gusakov:2008hv,Ofengeim:2019fjy}.

We are now left with four independent variables, the hyperon number $n_H$, baryon number $n_B$, and the electron and muon fraction $x_e=n_e/n_B$ and $x_{\mu}=n_{\mu}/n_B$. The continuity equations for baryon number, electron and muon number are given by 
\begin{align}
\frac{\partial n_{B}}{\partial t}-n_{B}\nabla\cdot\boldsymbol{v} & =0\,,\\
\frac{\partial n_{e}}{\partial t}-n_{e}\nabla\cdot\boldsymbol{v} & =0\,,\\
\frac{\partial n_{\mu}}{\partial t}-n_{\mu}\nabla\cdot\boldsymbol{v} & =0\,.
\end{align}
The hyperon number is not conserved but changes due to weak interactions. We assume that semi-leptonic Urca-type processes \cite{Alford:2019qtm} and purely leptonic processes~\cite{Alford:2010jf} are very slow compared to the density oscillation timescale, so the electron and muon fractions never deviate from their equilibrium values: $\delta x_{e}=\delta x_{\mu}=0$.

Additionally, we ignore processes that change the hyperon number by more than one unit and assume that they are slow compared to the ones that only change it by one unit. The source terms in the hyperon evolution equation are due to the four strangeness changing processes in Eqs.~(\ref{eq:allpr}). The four reactions lead to source terms 
\begin{equation}
\Delta\Gamma_{i}=\lambda_{i}\delta\mu_{i} \quad \mathrm{with} \quad i=\text{I},\ldots,\text{IV} \, ,
\end{equation}
where $\Delta\Gamma$ are the differences between the back and forward rates $\Gamma$ from Eq.~(\ref{eq:rateint}) which try to re-establish chemical equilibrium. We have assumed that the oscillation amplitude $\delta n$ is small enough so that $\delta\mu_i\ll T$, corresponding to the subthermal regime, where the linear approximation is valid.
Taking into account that $\delta\mu_{n}+\delta\mu_{\Lambda}=\delta\mu_{p}+\delta\mu_{\Sigma^{-}}$
due to the strong reaction from Eq.~(\ref{eq:fast}), we find that all processes equilibrate the same chemical potential, $\delta\mu_{\text{\sliver I}}=\delta\mu_{\text{\sliver II}}=\delta\mu_{\text{\sliver III}}=\delta\mu_{\text{\sliver IV}}\equiv\delta\mu$
with 
\begin{equation}\label{eq:muD}
\delta\mu=2\mu_{n}-\mu_{p}-\mu_{\Sigma^{-}}=\mu_{n}-\mu_{\Lambda} \, .
\end{equation}
Therefore, the hyperonic evolution equation is given by
\begin{equation}
\frac{\partial n_{H}}{\partial t}-\nabla\cdot(n_{H}\boldsymbol{v})=\Gamma_{\text{I}}+\Gamma_{\text{II}}+\Gamma_{\text{III}}+\Gamma_{\text{IV}}\,.
\end{equation}

The pressure $P$ and the chemical imbalance $\delta\mu$, which are
functions of $n_{B},\,n_{H},\,x_{e}$ and $x_{\mu}$, can be expanded
around equilibrium as
\begin{align}
P & =P_{0}+\delta P\,,\label{eq:PmP0}\\
\delta P & =\frac{\partial P}{\partial n_{B}}|_{x_{e},x_{\mu},n_{H},T}\delta n_{B}+\frac{\partial P}{\delta n_{H}}|_{x_{e},x_{\mu},n_{B},T}\delta n_{H}\,,
\end{align}
and 
\[
\delta\mu=\frac{\partial\delta\mu}{\partial n_{B}}|_{x_{e},x_{\mu},n_{H},T}\delta n_{B}+\frac{\partial\delta\mu}{\delta n_{H}}|_{x_{e},x_{\mu},n_{B},T}\delta n_{H}\,,
\]
where we used that $\delta x_e=\delta x_{\mu}=0$. Inserting the harmonic density oscillation into the continuity equations yields
\begin{align}
\delta n_{B} & =-\frac{n_{B}}{i\omega}\nabla\cdot\boldsymbol{v}\label{eq:dnb}\,,\\
\delta n_{H} & =-\frac{1}{i\omega}\left[n_{H}\nabla\cdot\boldsymbol{v}-\lambda\delta\mu\right]\label{eq:dnh}\,,
\end{align}
where $\lambda=\lambda_{\text{I}}+\lambda_{\text{II}}+\lambda_{\text{III}}+\lambda_{\text{IV}}$.
By inserting the expression for $\delta\mu$ into Eq.~(\ref{eq:dnh}) we find
\begin{align}
\delta n_{H} & =-\frac{\nabla\cdot\boldsymbol{v}}{i\omega}\left[n_{H}+\lambda\frac{n_{B}}{i\omega}\frac{\partial\delta\mu}{\partial n_{B}}\right]\left(1+\frac{i\lambda}{\omega n_{B}}\frac{\partial\delta\mu}{\delta x_{H}}\right)^{-1} \label{eq:dnhII}\,.
\end{align}
We now use Eq.~(\ref{eq:PmP0}), and 
Eqs.~(\ref{eq:dnhII}) and (\ref{eq:dnb}) for the perturbations $\delta n_H$  and $\delta n_B$,
to compute $P-P_0$, from which we can obtain the (real part of) the bulk viscosity as the coefficient of $\nabla\cdot\boldsymbol{v}$,
\begin{equation}\label{eq:bulk}
\Re\zeta=\frac{\lambda\dfrac{\partial P}{\partial x_{H}}\dfrac{\partial\delta\mu}{\partial n_{B}}}{\omega^{2}+\dfrac{\lambda^{2}}{n_{B}^{2}}\left(\dfrac{\partial\delta\mu}{\partial x_{H}}\right)^{\!\! 2}}=n_{B}\left(\frac{\partial\delta\mu}{\partial x_H}\right)^{-1}\frac{\partial\delta\mu}{\partial n_B}\frac{\partial P}{\partial x_{H}}\frac{\gamma}{\omega^{2}+\gamma^{2}}\,,
\end{equation}
where we have defined 
\begin{equation}\label{eq:gamma}
\gamma=B\lambda \qquad \mathrm{with\ the\ susceptibility }\qquad B\equiv\frac{1}{n_{B}}\frac{\partial\delta\mu}{\partial x_{H}}|_{x_{e},x_{\mu},n_{B},T}\,.
\end{equation} The various derivatives with respect to the hyperon and baryon number of the pressure and $\delta\mu$ are computed numerically from the EOS by changing the hyperon or baryon number while holding the other variables constant and solving the RMF equations. 
\section{Results and Discussion}\label{sec:results}
\subsection{Rates in Fermi Surface Approximation and Full Phase Space Calculation}
\begin{figure} [t]
\begin{center}
\hbox{\includegraphics[width=0.5\textwidth]{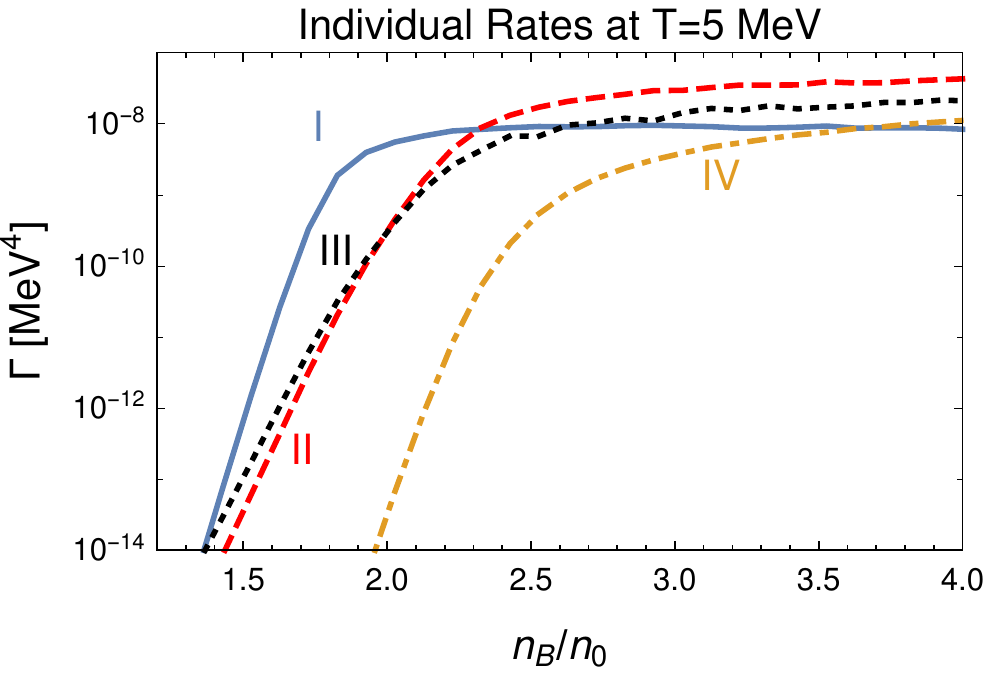}\includegraphics[width=0.5\textwidth]{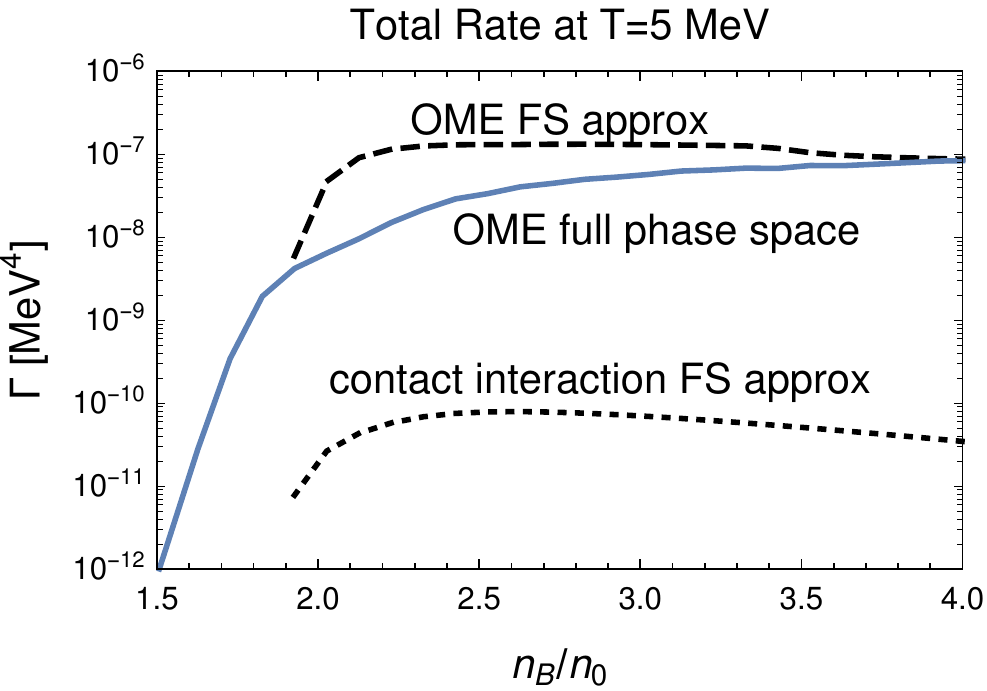}}
\caption{
{\it Left panel:} Individual rates of all four strangeness-changing processes defined in Eqs.~(\ref{eq:allpr}) at $T=5$\,MeV as a function of baryon density normalized to saturation density $n_0$ in the OME channel. Below the hyperon onset at  $n_B\approx 1.85\ n_0$, the rates drop quickly to zero as the thermal population of hyperons becomes highly suppressed.   {\it Right panel:} Sum of all rates at $T=5$\,MeV (solid blue line) in comparison to the FS approximation (black, dashed line labeled ``OME FS approx'') and the FS approximated rate of the contact interaction (black, dotted line) in the non-relativistic limit presented in Eqs.~(\ref{eq:MnnpsC}) and (\ref{eq:MnpplC}). Note that restricting to contact interactions in the ultra non-relativistic approximation means ignoring processes III and IV, which underestimates the total rate by a factor of $10^3$. Rates in the FS approximation are only defined above the hyperon density threshold.}
\label{fig:rates}
\end{center}
\end{figure}

In Fig.~\ref{fig:rates} we present our calculation of the rates for the processes I to IV in Eqs.~(\ref{eq:allpr}) as a function of baryon density at a temperature of $T=5$\,MeV. The left panel shows the four individual rates computed numerically from Eq.~(\ref{eq:rateint}). After analytical simplifications we carry out the remaining five dimensional integration using the CUBA library   \cite{HAHN200578}. At vanishing temperature, $T=0$, there exists a critical minimal density below which no hyperons are present. The actual value of the critical density highly depends on the choice of equation of state. For the PK1+H EOS that we are using, the onset density for hyperons at $T=0$ is $n_B\approx1.85\ n_0$. At non-zero temperature there is a thermal population of hyperons at and below this density. The thermal hyperon population increases with temperature and decreases when the density is lowered further. In this regime the hyperon density is exponentially sensitive to temperature and density, so we observe that the rates span many orders of magnitude. They are much less sensitive at densities above the hyperon threshold. We also observe that the rate of process IV is, especially at low densities, suppressed compared to all other rates. This is because this process involves three hyperons and only one nucleon, and therefore has less phase space available. Furthermore, the strong interaction in this case is mediated by kaon exchange instead of pion exchange, so the interaction is suppressed by the higher mass of the kaon $m_K$ in the meson propagator, even at high densities where the density of $\Lambda$ hyperons becomes comparable to the neutron density (see Fig.~\ref{fig:fracs}). 

\begin{figure} [t]
\begin{center}
\includegraphics[width=0.5\textwidth]{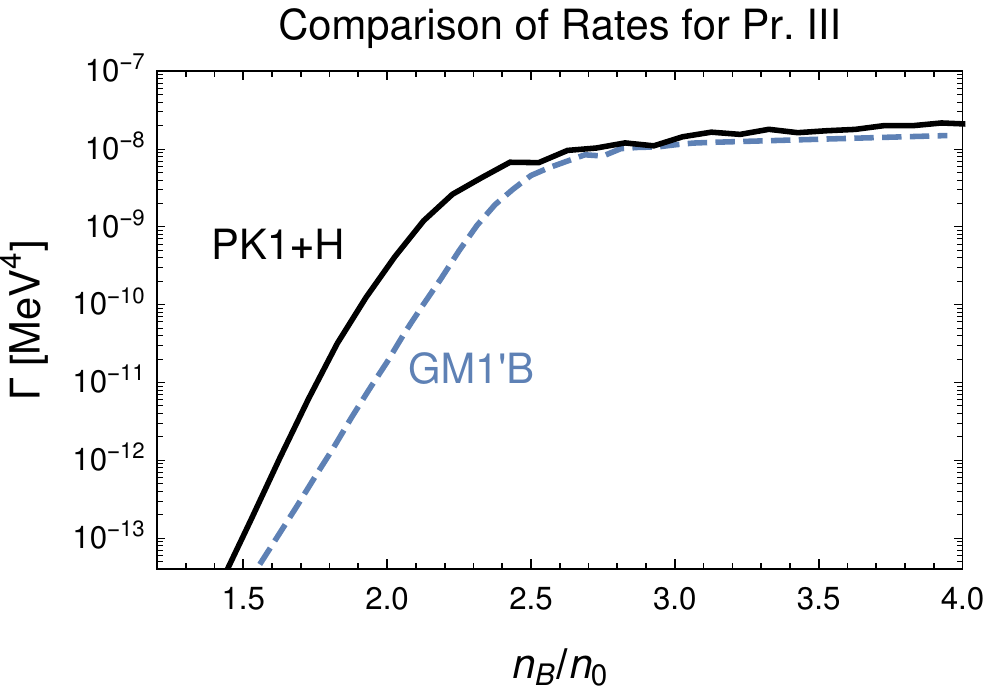}
\caption{ Comparison of the rate for process III (the dominant process at low densities) for the PK1+H EOS, which we use in in this work, and the GM1'B EOS (see Sec.~\ref{sec:hyperonic}) at $T=5$ MeV. The rates show very similar behavior, with GM1'B shifted slightly because it has a higher zero-temperature onset density for hyperons. }
\label{fig:eoscomp}
\end{center}
\end{figure}

It is interesting to compare these features with the GM1'B  EOS
which we have noted above is less convenient to deal with than PK1+H but is more consistent with phenomenological constraints.
In GM1'B the zero-temperature hyperon onset involves the $\Lambda$ rather than the $\Sigma^-$, and occurs at a higher density,  $n_B=2.39\, n_0$. However, the rates for processes II, III, and IV for the GM1'B EOS show very similar behavior to PK1+H, just shifted to slightly higher densities. Process I, since it involves the $\Sigma^-$, only occurs at higher densities. A direct comparison of the rates for process III, which is the dominant process at low densities, is shown in Fig.~\ref{fig:eoscomp} for a temperature of $T=5$ MeV. For our purposes, the important point is that both EOSes show the same pattern in their strangeless equilibration rate. We therefore expect that our results for PK1+H at densities close to hyperon onset are representative of hyperonic EOSes in general, close to their hyperon onset densities.

The right panel in Fig.~\ref{fig:rates} shows the total rate $\Gamma=\Gamma_\text{\text{I}}+\Gamma_\text{II}+\Gamma_\text{III}+\Gamma_\text{IV}$ at $T=5$\,MeV as a function of baryon density (solid blue line). For comparison, we show the total rate in the FS approximation in the OME channel (labeled OME FS approx) and for the sum of process I and II in the contact interaction channel. In the contact interaction channel, the ultra non-relativistic approximation has been performed, see paragraph below Eq.~(\ref{eq:MnpplC}) for more details.
All rates in the FS approximation are only computed above the hyperon threshold, since the Fermi momenta are not properly defined below the that density. At high densities, where the temperature becomes negligible compared to the Fermi momenta of the participating particles, the FS approximation works well for the OME contribution. However, it completely fails below the hyperon onset and overestimates the rate above the onset by up to an order of magnitude. Contrary to what one might expect, the FS approximation in the OME channel gives a faster rate than the full phase space integral, although the latter receives contributions from the thermally blurred Fermi surface and not only from particles exactly on their FS. However, further approximations in the energy integral tend to overestimate the rate, see App.~B of Ref.~\cite{Harris:2020qim} for more information.

\subsection{Re-equilibration Rates $\gamma$}

\begin{figure} [t]
\begin{center}
\hbox{\includegraphics[width=0.5\textwidth]{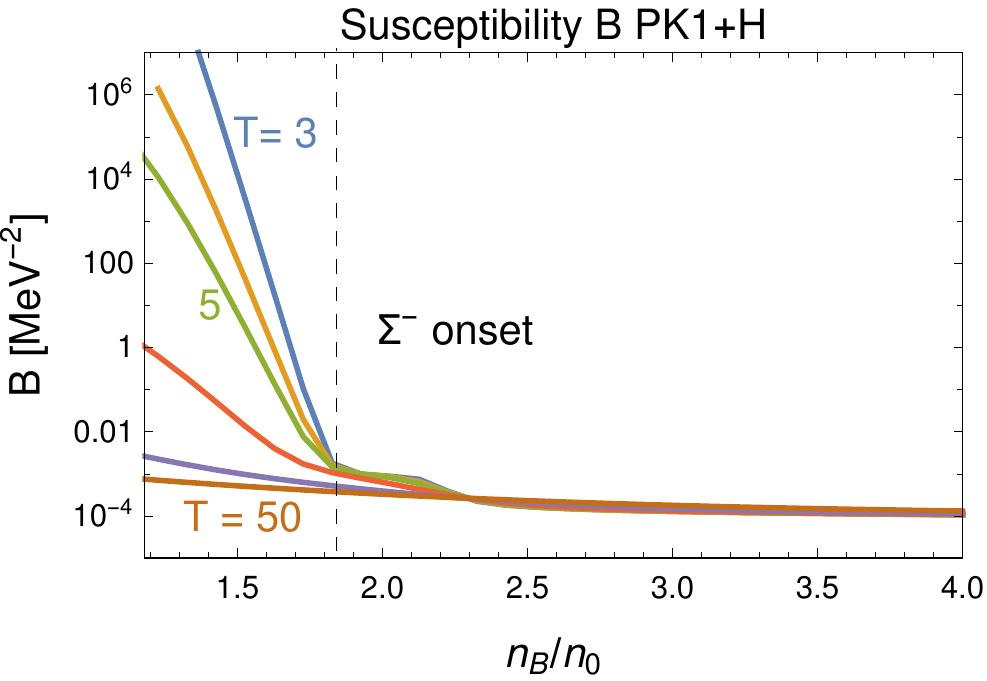}\hfill\includegraphics[width=0.5\textwidth]{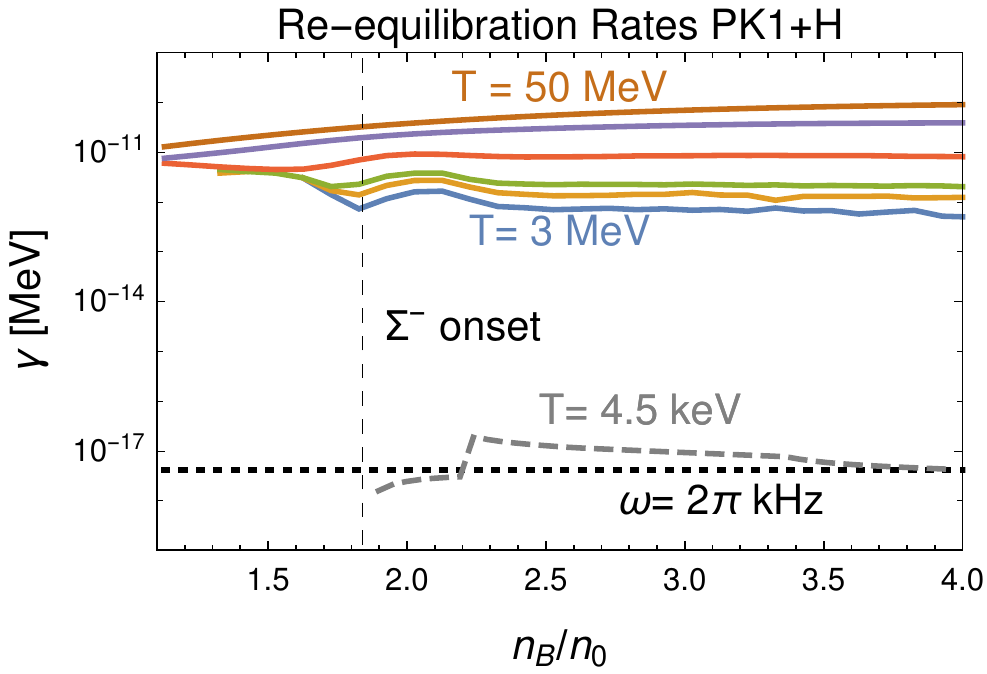}}
\caption{
\textit{Left panel: } Susceptibility $B$ for the PK1+H EOS (defined in Eq.~(\ref{eq:gamma})) as a function of baryon density at temperatures \hbox{$T = 3,\,4,\,5,\,10,\,25,\,50$\,MeV,} where the highest (blue solid) curve at low densities corresponds to $T=3$\,MeV and smaller susceptibilities correspond to higher temperatures. As the density drops below the hyperon onset at $n_B\approx 1.85\ n_0$  (marked by the thin, black, dashed, vertical line), the susceptibility shows an exponential increase, arising from the exponential suppression of the hyperon fraction in this regime. \textit{Right panel:} Re-equilibration rate $\gamma$ defined in Eq.~(\ref{eq:gamma}) for the PK1+H EOS as a function of baryon density. The color coding is identical to the left panel. All rates are obtained by evaluating the full phase space integral for the OME matrix elements. The black, dotted, horizontal line marks the optimal equilibration rate for maximal bulk viscosity where it would match the external oscillation, $\gamma=\omega$. The  exponential behavior of the rates from Fig.~\ref{fig:rates} balances that of the susceptibility $B$ from the left panel, leading to nearly density-independent re-equilibration rates $\gamma$. Therefore, the rates are, even at lower temperatures, too fast to match the external oscillation. Only at temperatures in the keV regime, $\gamma$ matches $\omega$, see the grey, dashed line where we show $\gamma$ for a temperature of $4.5\,$keV computed from the EOS at $T=0$ using the FS approximation for the rates. }
\label{fig:gamma}
\end{center}
\end{figure}
The chemical re-equilibration rate $\gamma$, defined in Eq.~(\ref{eq:gamma}), depends on the strangeness-changing rates and the susceptibility $B$, defined in Eq.~(\ref{eq:gamma}), which is proportional to the derivative of the chemical imbalance $\delta\mu$ with respect to the hyperon fraction $x_H$. We plot this susceptibility, for the PK1+H EOS, in the left panel of Fig.~\ref{fig:gamma} as a function of baryon density for temperatures from $T=3$ to $T=50$ MeV. As the density drops below the hyperon onset at a baryon density of $n_B\approx 1.85\ n_0$, the susceptibility shows an exponential increase. This can be understood in terms of the exponential density dependence of the thermal hyperon population (see Fig.~\ref{fig:fracs}). In this regime, the size of $\delta\mu$ necessary to change the strangeness by a given amount therefore also increases exponentially as the baryon density drops through the hyperon onset region. Above the hyperon onset, the strangeness fraction rises more slowly with increasing density, which leads to a leveling of $B$. For higher temperatures, the hyperon fraction and therefore the susceptibilities behave more smoothly. Combining these results with the rates shown in Fig.~\ref{fig:rates} allows us to compute the re-equilibration rate $\gamma$, which we show in the right panel of Fig.~\ref{fig:gamma}.
The opposite exponential density dependencies of the rates $\lambda$ and the suceptibility $B$ turn out to balance each other, so the re-equilibration rates do not
change significantly when densities drop below the hyperon onset. The horizontal black dotted line shows where the equilibration rate would match the external frequency $\omega$, which is where the bulk viscosity would reach its resonant maximum. In all our calculations we assume an external oscillation frequency of $\omega=2\pi$\,kHz which is typical for the high-amplitude density oscillations that occur immediately after the merger  \cite{Alford:2017rxf}.

For densities above saturation density (below which nuclear matter might not be uniform \cite{Chamel:2008ca}), and  temperatures down to about $2$ MeV, the equilbration rate remains far above the external oscillation frequency.
This leads us to expect that at the typical densities and temperatures of nuclear matter in neutron star mergers the hyperonic bulk viscosity and the resultant attenuation of density oscillations will not be significant. 
We also performed calculations at much lower temperatures (grey dashed lines), where
the FS approximation for the OME interaction yields rather accurate results.
Our calculations neglected hyperon and nuclear superfluidity which might become important at these temperatures \cite{Haensel:2001em,Ding:2016oxp,Sedrakian:2018ydt}. In this regime the difference between the hyperonic rates out of equilibrium can be computed by calculating the rate from Eq.~(\ref{eq:rateFS}) while linearizing $I(\xi)$ for small $\xi$, see Ref.~\cite{vanDalen:2003uy}. We computed the susceptibilities from the PK1+H EOS at $T=0$ and found that the bulk viscosity for 1\,kHz oscillations reaches a resonant maximum (neglecting superfluidity) at $T\approx 4$\,keV, which is in agreement with the findings of Ref.~\cite{Ofengeim:2019fjy}. 
We show this in Fig.~\ref{fig:gamma} and Fig.~\ref{fig:bulk} with a grey, dashed line. The kink in the re-equilibration rate arises from the sudden onset of the $\Lambda$-hyperon at $n_B\approx2.2\,n_0$ and is less pronounced in the higher temperature calculation, where the thermal population of hyperons blurs the onset.

\subsection{Bulk Viscosity and Dissipation Times}
\begin{figure} [t]
\begin{center}
\includegraphics[width=0.5\textwidth]{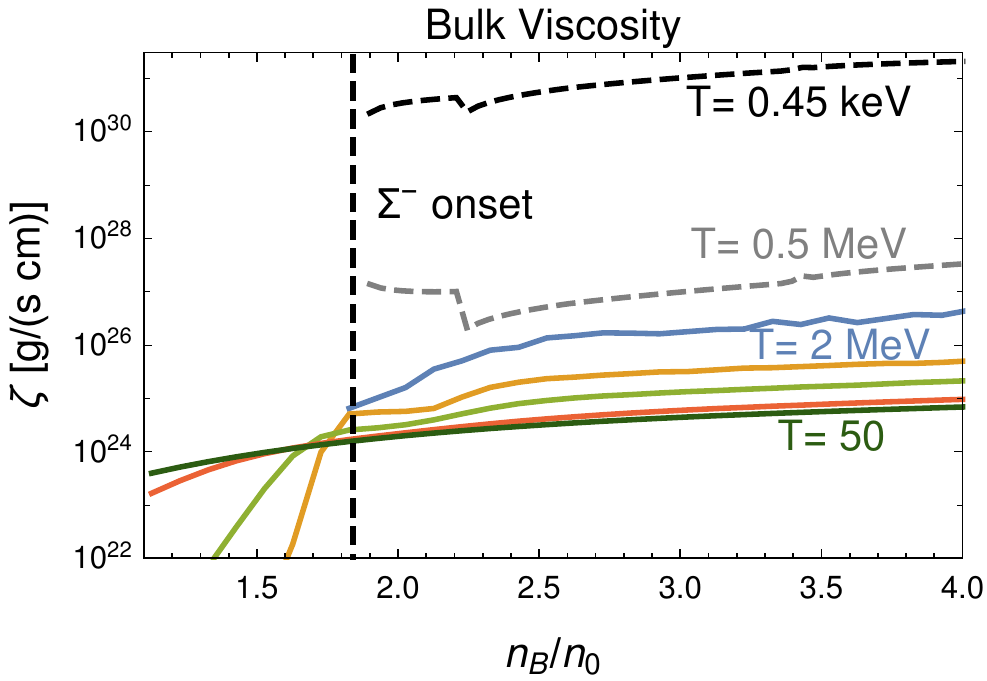}
\caption{Solid lines show the bulk viscosity as function of baryon density at temperatures  of $T= 2,\,5,\,10,\,25,\,50$\,MeV for the finite temperature PK1+H EOS computed from the full phase space integral. Above the hyperon onset, the bulk viscosity decreases with temperature, since even for $T=2$ MeV the re-equilibration rate, which generally further increases with temperature, is too fast to match the external frequency (see Fig.~\ref{fig:gamma}). Below the hyperon onset (marked by a black dashed vertical line), the bulk viscosity drops more drastically for low temperatures due to the faster decrease of the hyperon fraction. For temperatures of $2\,$MeV or below, the rates and susceptibilities can not be computed reliably below the hyperon onset because of the small hyperon fraction. At temperatures of over $20$ MeV, bulk viscosity is completely smooth due to the higher thermal hyperon population. The dashed lines show the bulk viscosity for $T=0.5\,$MeV and $T=4.5$\,keV. At temperatures of a few keV, the re-equilibration time and the external oscillation match, leading to a maximal bulk viscosity. The kink at $n_B\approx2.2\,n_0$ is a result of the $\Lambda$ onset. } 
\label{fig:bulk}
\end{center}
\end{figure}
\begin{figure} [t]
\begin{center}
\includegraphics[width=0.5\textwidth]{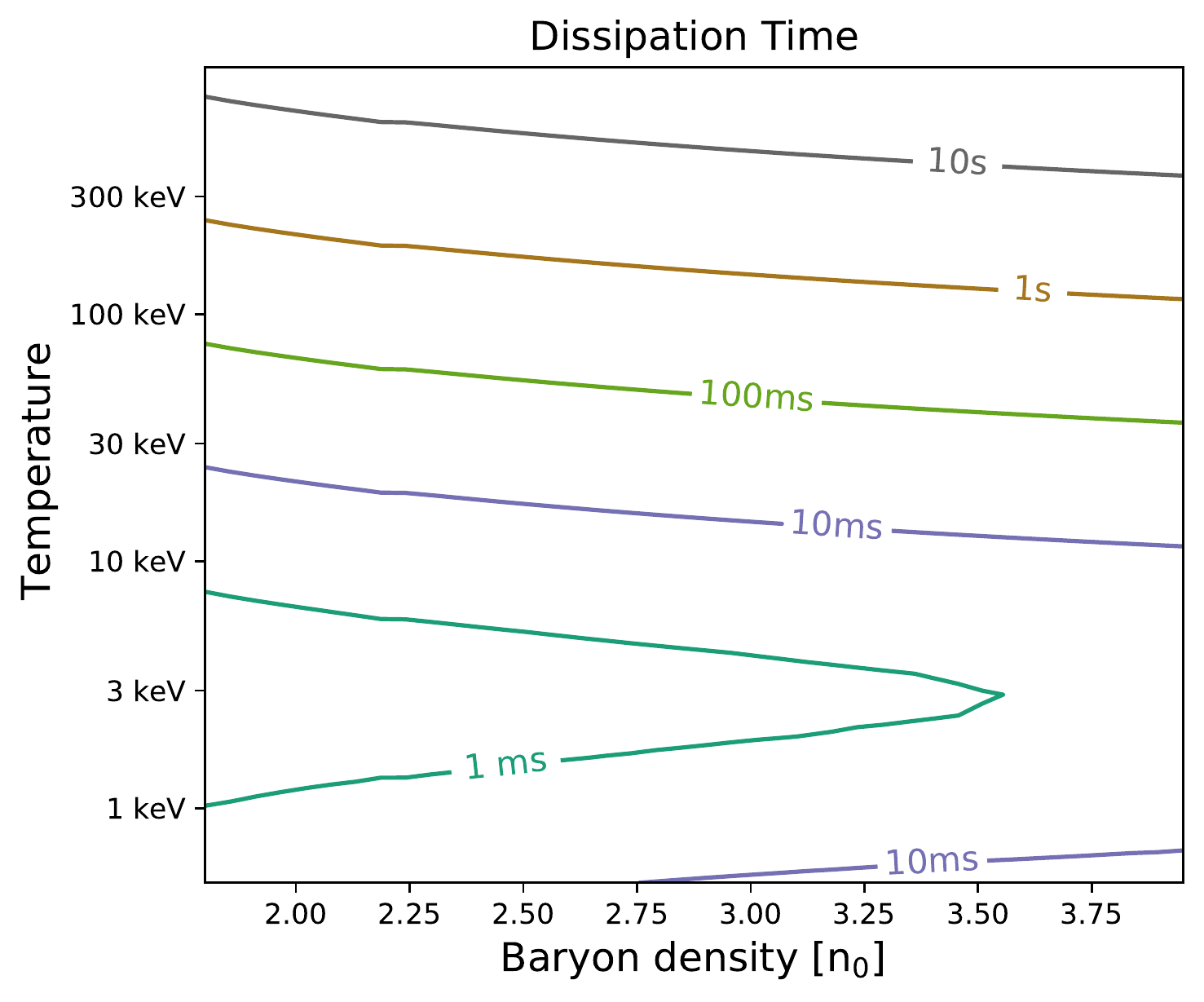}
\caption{Density and temperature dependence of the dissipation time for density oscillations, using the PK1+H EOS. We observe minimal dissipation times at a temperature of $T=4\,$keV, where the re-equilibration time $\gamma$ matches the external frequency $\omega$, leading to maximal bulk viscosity. For lower temperatures the re-equilibration rate is slower than the external oscillation, whereas for higher temperatures  the rates are too fast.
}
\label{fig:disslowT}
\end{center}
\end{figure}

In Fig.~\ref{fig:bulk} we show the bulk viscosity as a function of density at various temperatures.
The solid, colored lines are from Eq.~(\ref{eq:bulk}). We calculated the OME contribution to the equilibration rate by numerically evaluating the full phase space integral Eq.~(\ref{eq:rateint}). We cover baryon densities $n_B \in [1.4\,n_0,4\,n_0]$ and temperatures ranging from $T=2$\,MeV to $T=50$\,MeV.  The vertical dashed line indicates the density at which hyperons would first appear when the temperature is zero. The various colored lines correspond to the temperatures of $2,\,5,\,10,\,25,$ and $50$\,MeV.

At all densities for these temperatures we observe a low bulk viscosity compared to the bulk viscosity obtained from nuclear semi-leptonic processes \cite{Alford:2019qtm}. Above the hyperon onset, bulk viscosity decreases with temperature. The largest bulk viscosity is therefore obtained at the lowest shown temperature, $T=2$\,MeV. This is because the equilibration rate is always too fast (faster than the typical density oscillation frequency $\sim 1$\,kHz), so to increase the bulk viscosity one must decrease the equilibration rate, e.g., by reducing the temperature or density. As the density is lowered below the hyperon onset, bulk viscosity drops off much faster for smaller temperatures, since the thermal population decreases more rapidly. For high temperatures, in the tens of MeV, bulk viscosity only drops slowly with decreasing density, even below the $T=0$ hyperon onset. Therefore, the bulk viscosity for densities below the onset is larger for higher temperatures. 

In order to achieve a resonant match between the equilibration rate and the assumed density oscillation frequency of 1\,kHz, we must lower the temperature to the keV range. For these temperatures, we compute the susceptibilities from the zero-temperature PK1+H EOS, and calculate the rates in the FS approximation (neglecting superfluidity). We find that bulk viscosity peaks, for a given density above the hyperon onset, at temperatures around $4\,$\,keV, which is in agreement with Ref.~\cite{Ofengeim:2019fjy}. We show these results in Fig.~\ref{fig:bulk} using dashed, grey lines. The kink in the bulk viscosity is a result of the onset of the $\Lambda$ hyperon at a baryon density of $n_B\approx2.2\,n_0$ and is more pronounced for smaller temperatures, where few thermal hyperons are present.

For oscillations in mergers, one important measure of the importance of bulk viscosity is the dissipation time $\tau_{\mathrm{diss}}$ 
which  quantifies how fast a density oscillation of a fluid element is damped.
Following Refs.~\cite{disstime,Alford:2010gw, Alford:2019qtm},
\be
\tau_{\mathrm{diss}}\equiv\frac{\varepsilon}{d\varepsilon/dt}=\frac{\kappa_S^{-1}}{\omega^2\zeta} \, ,
\ee
where $\varepsilon$ is
the energy carried by an oscillation in baryon density with frequency $\omega$ and amplitude $\delta n$,
\be
\varepsilon=\frac{\kappa_S^{-1}}{2}\left(\frac{\delta n}{n_B}\right)^2 \, , \quad \mathrm{where} \quad \kappa_S^{-1}=n_B\frac{\partial P}{\partial n_B}|_{T,x_H,x_e,x_\mu} \, ,
\ee
and $\kappa_S$ is the incompressibility. 
The dissipation times we compute from the bulk viscosity for typical merger  temperatures in the MeV range is on the scale of seconds and above. Dissipation times at and around the resonant temperature of a few keV are considerably shorter, as can be seen in Fig.~\ref{fig:disslowT}, which shows a contour plot of the dissipation time in the plane of baryon density and temperature. In this plot, the resonant nature of the bulk viscosity is clearly visible: At a given density, the re-equilibration rate for temperatures below $1\,$ keV is too slow to match the external frequency, leading to a smaller bulk viscosity and longer dissipation times. Raising the temperature leads to a resonant maximum of the bulk viscosity and a corresponding minimum of the dissipation times, then at higher temperatures the rates become too fast and the dissipation times rise again. The density dependence of the dissipation time is much weaker than the $T^3$ temperature dependence of the rates $\Gamma$.

\section{Conclusions}
In this paper we have presented a calculation of hyperonic bulk viscosity and the resultant dissipation time for density oscillations in the range of densities and temperatures that are expected to exist in binary neutron star mergers. For this purpose, we used the PK1+H EOS, whose maximum neutron star mass is at the edge of compatibility with observations, but we checked that comparable results would be obtained for the GM1'B EOS: for both these EOS the hyperonic bulk viscosity is small compared to its nuclear counterpart in the MeV temperature range.

We calculated hyperonic equilibration rates by evaluating the one meson exchange contribution, which, as first discussed in Ref.~\cite{vanDalen:2003uy}, is the dominant channel in all of the studied parameter space. 

Previous studies of hyperonic bulk viscosity used the Fermi Surface approximation, since they were concerned with temperatures in the keV range. The typical temperature in mergers is in the MeV range, which is high enough to invalidate the FS approximation. We therefore numerically evaluate the full phase space integral for the rates. This allows us to study the behavior of the system at densities below the zero-temperature hyperon onset, where there is only a thermal population of hyperons and the Fermi surface is not well defined.  We find that at temperatures $T\gtrsim 1$\,MeV the hyperonic bulk viscosity for kHz density oscillations is always much smaller than its nuclear counterpart \cite{Alford:2019qtm,Alford:2019kdw}. This is because the
beta re-equilibration rate 
is always too fast to match an external frequency oscillation of $\omega\sim 2\pi$\,kHz. 

Consequently, it seems that hyperonic bulk viscosity is not a significant source of damping of density oscillations in neutron star mergers. 

In future work on viscosity, the influence of large amplitude oscillations and magnetic fields on the hyperon bulk viscosity could be studied.  Above a temperature of roughly $5$\,MeV, neutrino trapping, which we have neglected in this treatment, would likely become important so beta equilibration processes with neutrinos in the initial state would have to be included. 

Finally, we note that hyperonic decays might play an important role in other transport phenomena, like radiative dissipation \cite{Sad:2009hba} or phase conversion dissipation \cite{Alford:2014jha}. The rate calculations presented in this paper are a necessary step towards extending these calculations to the higher temperatures that occur in mergers. We also computed the dissipation times at keV temperatures, where the bulk viscosity reaches its resonant maximum. For temperatures around $T\approx4$\,keV we find dissipation times of a few ms. This suggests that hyperonic bulk viscosity might play an important role in the damping of induced oscillations in highly eccentric neutron star mergers, where temperatures are much lower than in the post-merger phase. 

\section{Acknowledgements}
We thank Lorenzo Andreoli, Mikhail Gusakov, Steven Harris, Andreas Schmitt and Ziyuan Zhang for useful discussions.
This research was partly supported by the U.S. Department of Energy, Office of Science, Office of Nuclear Physics, under Award No.\ \#DE-FG02-05ER41375.

\appendix
\section{Feynman and Quarkflow Diagrams}
In this appendix we present the Feynman diagrams and the corresponding quark flow diagrams for three of the four strangeness changing processes we take into account, see Eqs.~(\ref{eq:allpr}). Process I is depicted in the main part of this publication, see Fig.~\ref{fig:_feyndia}. For the computation of the matrix element in Eq.~(\ref{eq:matelem}), a second Feynman diagram with the initial baryons exchanged has to be subtracted. Only for process II this leads to a nontrivial change, since in all other cases the initial particles are identical. In these trivial cases, we do not draw the second Feynman and quark flow diagram.

For process I and II, we additionally show the diagrams for the same process in the contact interaction channel, where the baryons directly exchange a charged $W$-boson. These diagrams are the basis for the matrix elements in Eq.~(\ref{eq:MnnpsC}) and Eq.~(\ref{eq:MnpplC}).
\label{app:dia}
\begin{figure}[h]
\begin{center}
\begin{tabular}{c@{\hspace{4em}}c@{\hspace{4em}}c}

\begin{tikzpicture}
\begin{feynhand}
\vertex [ringdot] (a) at (0,0) {} ;
\vertex [dot] (b) at (0,-2) {};
\node at (0.1,0.4) {F$^W_{n\Lambda}$};
\node at (0.1,-2.4) {F$^S_{pp}$};
\vertex [particle] (c) at (-2,0) {n};
\vertex [particle] (d) at (2,0) {$\Lambda$};
\vertex [particle] (i3) at (-2,-2) {p};
\vertex [particle] (i4) at (2,-2) {p};
\propagator[plain] (c) to (a);
\propagator[plain] (a) to (d);
\propagator [scalar] (a) to [edge label = $\pi_0$](b);
\propagator[plain] (i3) to (b);
\propagator[plain] (b) to (i4);
\end{feynhand}
\end{tikzpicture}
&
\begin{tikzpicture}[scale=.5]
\begin{feynhand}
\vertex [particle] (o1) at (4,0) ;
\vertex [particle] (o2) at (4,-1) ;
\vertex [ringdot] (W1) at (3,-1) ;
\vertex [ringdot] (W2) at (3,-2) ;
\vertex [dot](o3) at (4,-2) ;
\vertex [dot] (u1) at (4,-4) ;
\vertex [particle](u2) at (4,-5) ;
\vertex [particle] (u3) at (4,-6) ;
\vertex [particle] (q1) at (0,0) {d};
\vertex [particle] (q2) at (0,-1) {u};
\vertex [particle] (q3) at (0,-2) {d};
\vertex [particle] (q4) at (0,-4) {u};
\vertex [particle] (q5) at (0,-5) {u};
\vertex [particle] (q6) at (0,-6) {d};

\vertex [particle] (q1p) at (8,0) {d};
\vertex [particle] (q2p) at (8,-1) {s};
\vertex [particle] (q3p) at (8,-2) {u};
\vertex [particle] (q4p) at (8,-4) {u};
\vertex [particle] (q5p) at (8,-5) {u};
\vertex [particle] (q6p) at (8,-6) {d};

\propagator[fermion] (q1) to (o1);
\propagator[fermion] (q2) to (W1);
\propagator[plain] (W1) to  [edge label =$s$] (o2);
\propagator[fermion] (q3) to  (W2);
\propagator[plain] (W2) to [edge label =$u$](o3);

\propagator[chabos] (W1) to [edge label'=$W^+$] (W2);

\propagator[fermion] (q4) to (u1);
\propagator[fermion] (q5) to (u2);
\propagator[fermion] (q6) to (u3);

\propagator[fermion] (o1) to (q1p);
\propagator[fermion] (o2) to (q2p);
\propagator[fermion] (o3) to (q3p);

\propagator[fermion] (u1) to (q4p);
\propagator[fermion] (u2) to (q5p);
\propagator[fermion] (u3) to (q6p);

\propagator[scalar] (o3) to (u1) ;

\node at (4.8,-3.0) {$\pi^0$};
\node at (-.8,-1) {n};
\node at (-.8,-5) {p};
\node at (8.8,-1) {$\Lambda$};
\node at (8.8,-5) {p};

\end{feynhand}
\end{tikzpicture}
&
\begin{tikzpicture}[scale=.5]
\begin{feynhand}

\vertex [particle] (lu) at (-2,-2) {p};
\vertex [particle] (ro) at (2,2) {$\Lambda$};
\vertex [particle] (lo) at (-2,2) {n};
\vertex [particle] (ru) at (2,-2) {p};
\propagator[plain] (lu) to (ro);
\propagator[plain] (ru) to (lo);

\end{feynhand}
\end{tikzpicture}
\\
(a) & (b) & (c)
\end{tabular}
\caption{Feynman- and quarkflow diagram for process II, $n+p\to p+\Lambda$, in the OME channel (panels (a) and (b)) and the contact interaction channel (panel (c)). }
\label{Fig:quarkIIa}
\end{center}
\end{figure}
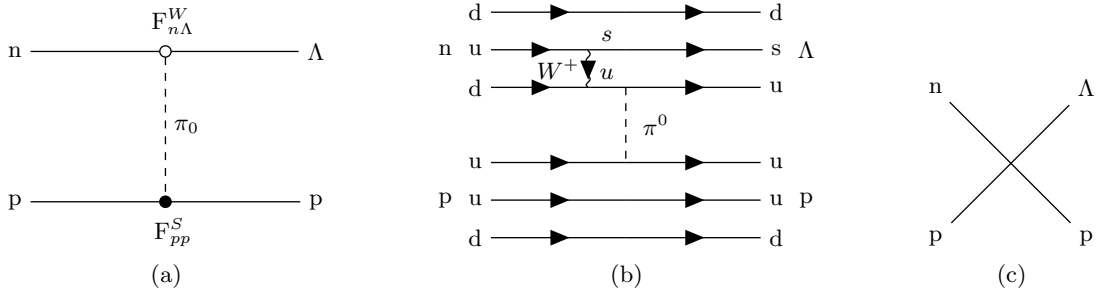
\begin{figure}[h]
\begin{center}
\begin{tabular}{c@{\hspace{4em}}c@{\hspace{4em}}c}
\begin{tikzpicture}
\begin{feynhand}
\vertex [ringdot] (a) at (0,0) {} ;
\vertex [dot] (b) at (0,-2) {};
\node at (0.1,0.4) {F$^W_{p\Lambda}$};
\node at (0.1,-2.4) {F$^S_{np}$};
\vertex [particle] (c) at (-2,0) {p};
\vertex [particle] (d) at (2,0) {$\Lambda$};
\vertex [particle] (i3) at (-2,-2) {n};
\vertex [particle] (i4) at (2,-2) {p};
\propagator[plain] (c) to (a);
\propagator[plain] (a) to (d);
\propagator [scalar] (a) to [edge label = $\pi^-$](b);
\propagator[plain] (i3) to (b);
\propagator[plain] (b) to (i4);
\end{feynhand}
\end{tikzpicture}
&
\begin{tikzpicture}[scale=.5]
\begin{feynhand}
\vertex [particle] (o1) at (4,0) ;
\vertex [particle] (o2) at (4,-1) ;
\vertex [ringdot] (W1) at (3,-1) ;
\vertex [ringdot] (W2) at (3,-2) ;
\vertex [dot](o3) at (4,-2) ;
\vertex [dot] (u1) at (4,-4) ;
\vertex [particle](u2) at (4,-5) ;
\vertex [particle] (u3) at (4,-6) ;
\vertex [particle] (q1) at (0,0) {u};
\vertex [particle] (q2) at (0,-1) {u};
\vertex [particle] (q3) at (0,-2) {d};
\vertex [particle] (q4) at (0,-4) {d};
\vertex [particle] (q5) at (0,-5) {d};
\vertex [particle] (q6) at (0,-6) {u};

\vertex [particle] (q1p) at (8,0) {u};
\vertex [particle] (q2p) at (8,-1) {s};
\vertex [particle] (q3p) at (8,-2) {d};
\vertex [particle] (q4p) at (8,-4) {u};
\vertex [particle] (q5p) at (8,-5) {d};
\vertex [particle] (q6p) at (8,-6) {u};

\propagator[fermion] (q1) to (o1);
\propagator[fermion] (q2) to (W1);
\propagator[plain] (W1) to  [edge label =$s$] (o2);
\propagator[fermion] (q3) to  (W2);
\propagator[plain] (W2) to [edge label =$u$](o3);

\propagator[chabos] (W1) to [edge label'=$W^+$] (W2);

\propagator[fermion] (q4) to (u1);
\propagator[fermion] (q5) to (u2);
\propagator[fermion] (q6) to (u3);

\propagator[fermion] (o1) to (q1p);
\propagator[fermion] (o2) to (q2p);
\propagator[fermion] (o3) to (q3p);

\propagator[fermion] (u1) to (q4p);
\propagator[fermion] (u2) to (q5p);
\propagator[fermion] (u3) to (q6p);

\propagator[anti charged scalar] (o3) to (u1) ;

\node at (5.5,-3.0) {$\pi^-=\bar{u}d$};
\node at (-.8,-1) {p};
\node at (-.8,-5) {n};
\node at (8.8,-1) {$\Lambda$};
\node at (8.8,-5) {p};

\node at (4.8,-1.6) {d};
\node at (4.8,-3.6) {u};

\end{feynhand}
\end{tikzpicture}
&
\begin{tikzpicture}[scale=.5]
\begin{feynhand}

\vertex [particle] (lu) at (-2,-2) {n};
\vertex [particle] (ro) at (2,2) {$\Lambda$};
\vertex [particle] (lo) at (-2,2) {p};
\vertex [particle] (ru) at (2,-2) {p};
\propagator[plain] (lu) to (ro);
\propagator[plain] (ru) to (lo);

\end{feynhand}
\end{tikzpicture}
\\
(a) & (b) & (c)
\end{tabular}
\caption{Feynman- and quarkflow diagram for process II, $n+p\to p+\Lambda$, in the OME channel (panels (a) and (b)) and the contact interaction channel (panel c), both with the initial nucleons exchanged. }
\label{Fig:quarkIIb}
\end{center}
\end{figure}
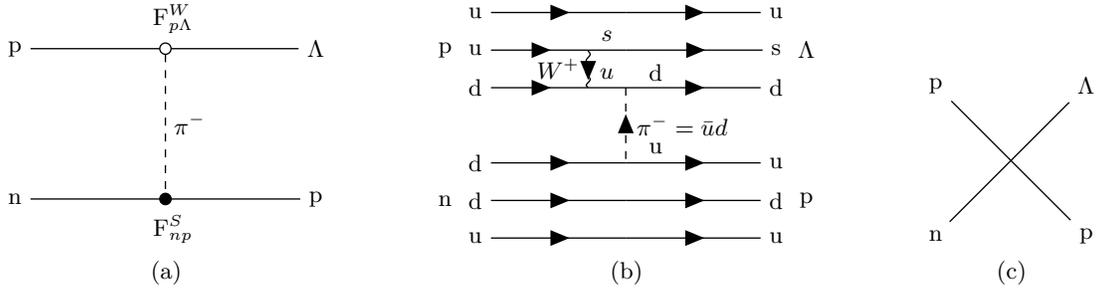
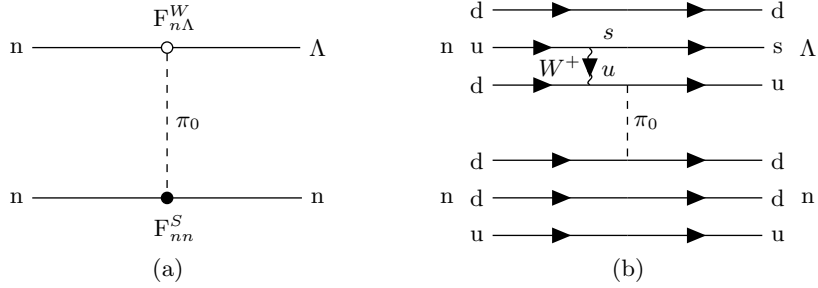
\begin{figure}[h]
\begin{center}
\begin{tabular}{c@{\hspace{4em}}c@{\hspace{4em}}}
\begin{tikzpicture}
\begin{feynhand}
\vertex [ringdot] (a) at (0,0) {} ;
\vertex [dot] (b) at (0,-2) {};
\node at (0.1,0.4) {F$^W_{n\Lambda}$};
\node at (0.1,-2.4) {F$^S_{nn}$};
\vertex [particle] (c) at (-2,0) {n};
\vertex [particle] (d) at (2,0) {$\Lambda$};
\vertex [particle] (i3) at (-2,-2) {n};
\vertex [particle] (i4) at (2,-2) {n};
\propagator[plain] (c) to (a);
\propagator[plain] (a) to (d);
\propagator [scalar] (a) to  [edge label = $\pi_0$](b);
\propagator[plain] (i3) to (b);
\propagator[plain] (b) to (i4);
\end{feynhand}
\end{tikzpicture}
&
\begin{tikzpicture}[scale=0.5]
\begin{feynhand}
\vertex [particle] (o1) at (4,0) ;
\vertex [particle] (o2) at (4,-1) ;
\vertex [ringdot] (W1) at (3,-1) ;
\vertex [ringdot] (W2) at (3,-2) ;
\vertex [dot](o3) at (4,-2) ;
\vertex [dot] (u1) at (4,-4) ;
\vertex [particle](u2) at (4,-5) ;
\vertex [particle] (u3) at (4,-6) ;
\vertex [particle] (q1) at (0,0) {d};
\vertex [particle] (q2) at (0,-1) {u};
\vertex [particle] (q3) at (0,-2) {d};
\vertex [particle] (q4) at (0,-4) {d};
\vertex [particle] (q5) at (0,-5) {d};
\vertex [particle] (q6) at (0,-6) {u};

\vertex [particle] (q1p) at (8,0) {d};
\vertex [particle] (q2p) at (8,-1) {s};
\vertex [particle] (q3p) at (8,-2) {u};
\vertex [particle] (q4p) at (8,-4) {d};
\vertex [particle] (q5p) at (8,-5) {d};
\vertex [particle] (q6p) at (8,-6) {u};

\propagator[fermion] (q1) to (o1);
\propagator[fermion] (q2) to (W1);
\propagator[plain] (W1) to  [edge label =$s$] (o2);
\propagator[fermion] (q3) to  (W2);
\propagator[plain] (W2) to [edge label =$u$](o3);

\propagator[chabos] (W1) to [edge label'=$W^+$] (W2);

\propagator[fermion] (q4) to (u1);
\propagator[fermion] (q5) to (u2);
\propagator[fermion] (q6) to (u3);

\propagator[fermion] (o1) to (q1p);
\propagator[fermion] (o2) to (q2p);
\propagator[fermion] (o3) to (q3p);

\propagator[fermion] (u1) to (q4p);
\propagator[fermion] (u2) to (q5p);
\propagator[fermion] (u3) to (q6p);

\propagator[scalar] (o3) to (u1) ;

\node at (4.5,-3.0) {$\pi_0$};
\node at (-.8,-1) {n};
\node at (-.8,-5) {n};
\node at (8.8,-1) {$\Lambda$};
\node at (8.8,-5) {n};

\end{feynhand}
\end{tikzpicture}
\\
(a)&(b)
\end{tabular}
\caption{Feynman- and quarkflow diagram for process III, $n+n\to n+\Lambda$ in the OME channel. The corresponding contact interaction channel would be mediated by neutral $Z-$boson exchange and is therefore suppressed by the GIM mechanism. For the calculation of the OME matrix element, a diagram with the two incoming neutrons exchanged has to be subtracted from the depicted one. }
\label{Fig:quarkIII}
\end{center}
\end{figure}

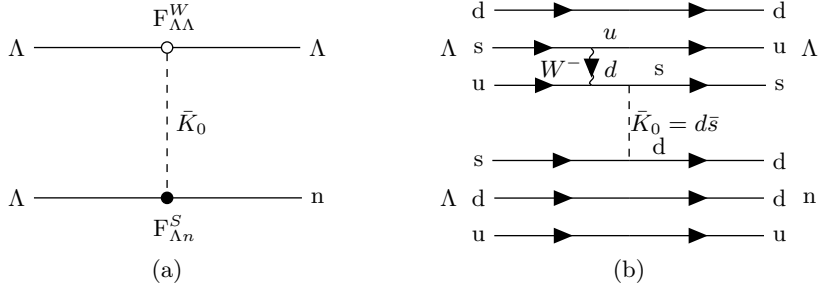
\begin{figure}[h]
\begin{center}
\begin{tabular}{c@{\hspace{4em}}c@{\hspace{4em}}}
\begin{tikzpicture}
\begin{feynhand}
\vertex [ringdot] (a) at (0,0) {} ;
\vertex [dot] (b) at (0,-2) {};
\node at (0.1,0.4) {F$^W_{\Lambda\Lambda}$};
\node at (0.1,-2.4) {F$^S_{\Lambda n}$};
\vertex [particle] (c) at (-2,0) {$\Lambda$};
\vertex [particle] (d) at (2,0) {$\Lambda$};
\vertex [particle] (i3) at (-2,-2) {$\Lambda$};
\vertex [particle] (i4) at (2,-2) {n};
\propagator[plain] (c) to (a);
\propagator[plain] (a) to (d);
\propagator [scalar] (a) to [edge label = $\bar{K}_0$](b);
\propagator[plain] (i3) to (b);
\propagator[plain] (b) to (i4);
\end{feynhand}
\end{tikzpicture}
&
\begin{tikzpicture}[scale=0.5]
\begin{feynhand}
\vertex [particle] (o1) at (4,0) ;
\vertex [particle] (o2) at (4,-1) ;
\vertex [ringdot] (W1) at (3,-1) ;
\vertex [ringdot] (W2) at (3,-2) ;
\vertex [dot](o3) at (4,-2) ;
\vertex [dot] (u1) at (4,-4) ;
\vertex [particle](u2) at (4,-5) ;
\vertex [particle] (u3) at (4,-6) ;
\vertex [particle] (q1) at (0,0) {d};
\vertex [particle] (q2) at (0,-1) {s};
\vertex [particle] (q3) at (0,-2) {u};
\vertex [particle] (q4) at (0,-4) {s};
\vertex [particle] (q5) at (0,-5) {d};
\vertex [particle] (q6) at (0,-6) {u};

\vertex [particle] (q1p) at (8,0) {d};
\vertex [particle] (q2p) at (8,-1) {u};
\vertex [particle] (q3p) at (8,-2) {s};
\vertex [particle] (q4p) at (8,-4) {d};
\vertex [particle] (q5p) at (8,-5) {d};
\vertex [particle] (q6p) at (8,-6) {u};

\propagator[fermion] (q1) to (o1);
\propagator[fermion] (q2) to (W1);
\propagator[plain] (W1) to  [edge label =$u$] (o2);
\propagator[fermion] (q3) to  (W2);
\propagator[plain] (W2) to [edge label =$d$](o3);

\propagator[chabos] (W1) to [edge label'=$W^-$] (W2);

\propagator[fermion] (q4) to (u1);
\propagator[fermion] (q5) to (u2);
\propagator[fermion] (q6) to (u3);

\propagator[fermion] (o1) to (q1p);
\propagator[fermion] (o2) to (q2p);
\propagator[fermion] (o3) to (q3p);

\propagator[fermion] (u1) to (q4p);
\propagator[fermion] (u2) to (q5p);
\propagator[fermion] (u3) to (q6p);

\propagator[scalar] (o3) to (u1) ;

\node at (5.2,-3.0) {$\bar{K}_0=d\bar{s}$};
\node at (-.8,-1) {$\Lambda$};
\node at (-.8,-5) {$\Lambda$};
\node at (8.8,-1) {$\Lambda$};
\node at (8.8,-5) {n};

\node at (4.8,-1.6) {s};
\node at (4.8,-3.6) {d};

\end{feynhand}
\end{tikzpicture}
\\
(a)&(b)
\end{tabular}
\caption{Feynman- and quarkflow diagram for process IV, $\Lambda+\Lambda\to \Lambda+n$ in the OME channel. The corresponding contact interaction channel is suppressed due to the GIM mechanism. For the calculation of the OME matrix element, a diagram with the two incoming hyperons exchanged has to be subtracted from the depicted one. }
\label{Fig:quarkIV}
\end{center}
\end{figure}
\clearpage

\section{Numerical Parameters and Coupling Constants}
\label{app:const}
In this appendix we collect all numerical parameters and coupling constants from the EOS and the Feynman diagrams in Fig.~\ref{fig:_feyndia} and App.~\ref{app:dia}.
\begin{table}[h]
\begin{tabular}{|c|c|c|c|c|c|c|c|c|c|c|}
\hline 
$M_{n}$ & $M_{p}$ & $m_{\sigma}$ & $m_{\omega}$ & $m_{\rho}$ & $g_{\sigma N}$ & $g_{\omega N}$ & $g_{\rho N}$ & $g_{2} [fm^{-1}]$ & $g_{3}$ & $c_{3}$\tabularnewline
\hline 
\hline 
939.5731 & 938.2796 & 514.0891 & 784.254 & 763 & 10.3222 & 13.0131 & 4.5297 & -8.1688 & -9.9976 & 55.636\tabularnewline
\hline 
\end{tabular}
\vspace*{4em}
\begin{tabular}{|c|c|c|c|c|c|c|c|c|c|c|c|}
\hline 
$m_e$& $m_{\mu}$&$m_\pi$&$m_K$ &$M_{\Lambda}$ & $M_{\Sigma^{-}}$ & $g_{\sigma\Lambda}$ & $g_{\sigma\Sigma^{-}}$ & $g_{\omega\Lambda}$ & $g_{\omega\Sigma^{-}}$ & $g_{\rho\Lambda}$ & $g_{\rho\Sigma^{-}}$\tabularnewline
\hline 
\hline 
$0$&$106$&$134.976$&$497.611$&$1115$  & $1197$  & $0.642$ $g_{\sigma N}$ & $0.453$ $g_{\sigma N}$ & $0.66\,g_{\omega N}$ & $0.66\,g_{\omega N}$ & $0$ & $-2\,g_{\rho N}$\tabularnewline
\hline 
\end{tabular}

\caption{Numerical parameters for the nuclear part and the hyperonic extension of the
PK1+H equation of state. The nuclear EOS and all parameters are taken from Ref.~\cite{Long:2003dn}. The meson-nucleon Yukawa couplings are identical for neutron and proton, i.e.\, $g_{\sigma N}\equiv g_{\sigma n}=g_{\sigma p}$ etc.. All masses are given in MeV.}

\end{table}

\begin{table}[h]
\begin{tabular}{|c|c|c|c|c|c|}
\hline 
Vertex & $g_{ij}$ & $A_{ij}$ & $B_{ij}$\tabularnewline
\hline 
\hline 
$pp\pi$ & $13.3$ & - & -\tabularnewline
\hline 
$np\pi$ & $13.3\sqrt{2}$ & - & -\tabularnewline
\hline 
$nn\pi$ & $-13.3$ & - & -\tabularnewline
\hline 
$\Lambda n\pi$ & - & $-1.07$ & $-7.19$\tabularnewline
\hline 
$\Lambda p\pi$ & - & $1.46$ & $9.95$\tabularnewline
\hline 
$\Sigma^- n\pi$ & - & $1.93$ & $-0.63$\tabularnewline
\hline 
$\Lambda nK$ & $-14.1$ & - & -\tabularnewline
\hline 
$\Lambda KK$ & - & $0.67$ & $-12.72$\tabularnewline
\hline 
\end{tabular}

\caption{Coupling constants for the matrix element in the OME channel taken from Refs.~\cite{vanDalen:2003uy,Ofengeim:2019fjy}. The kaon couplings were originally published in Refs.~\cite{Parreno:1996if,Parreno:2001xv}. The vertices are defined in Eq.~(\ref{eq:vertex}).}

\end{table}

\begin{table}[h]
\begin{tabular}{|c|c|c|c|}
\hline 
$c_A^{np}$ & $c_A^{p\Lambda}$ & $c_A^{n\Sigma^-}$ & $\sin^2(2\theta_c)$ 
\tabularnewline
\hline 
\hline 
-1.26 & -0.72 & 0.34 & 0.18742 \tabularnewline
\hline 
\end{tabular}

\caption{Coupling constants for the matrix element in the contact interaction channel taken from Refs.~\cite{Lindblom:2001hd,vanDalen:2003uy}}

\end{table}
\clearpage

\bibliography{hyperon_bulkvis}

\end{document}